\documentclass[useAMS,usenatbib]{mnras}

\usepackage{times}
\usepackage{color}
\usepackage{breakurl}
\usepackage{color}
\usepackage{mathrsfs}
\usepackage{multicol}
\usepackage{bm}		
\usepackage{pdflscape}	
\usepackage{amsfonts,amsmath,amssymb,mathrsfs}
\usepackage{graphicx,epsf,epsfig}
\usepackage{inputenc}
\usepackage{subfigure}
\usepackage{bm}
\usepackage{longtable}
\usepackage[usenames,dvipsnames]{xcolor}
\usepackage{multirow}
\usepackage{hyperref}
\usepackage{times}
\usepackage{color}
\usepackage{tabularx}
\usepackage{appendix}
\usepackage{epstopdf}

\def\be{\begin{equation}}
\def\ee{\end{equation}}
\def\bea{\begin{eqnarray}}
\def\eea{\end{eqnarray}}

\title[]{Constraints on the transition redshift from the calibrated Gamma-ray Burst $E_{\rm p}$--$E_{\rm iso}$ correlation}

\author[]{Marco Muccino$^{1}$\thanks{marco.muccino@lnf.infn.it},  Orlando Luongo$^{1,2,3}$\thanks{orlando.luongo@unicam.it}, Deepak Jain$^{4}$\thanks{djain@ddu.du.ac.in}\\
$^{1}$NNLOT, Al-Farabi Kazakh National University, Al-Farabi av. 71,  Almaty, 050040, Kazakhstan.\\
$^{2}$Scuola di Scienze e Tecnologie, Universit\`a di Camerino, Via Madonna delle Carceri 9,  Camerino, 62032, Italy.\\
$^{3}$Dipartimento di Matematica, Universit\`a di Pisa, Largo Bruno Pontecorvo 5, Pisa, 56127, Italy.\\
$^{4}$Deen Dayal Upadhyaya College, University of Delhi, Dwarka, New Delhi 110078, India.
}

\date{Accepted XXX. Received YYY; in original form ZZZ}

\pubyear{2022}

\begin{document}
\label{firstpage}
\pagerange{\pageref{firstpage}--\pageref{lastpage}}
\maketitle

\begin{abstract}
We constrain the deceleration-acceleration epoch, namely the transition redshift $z_{tr}$, adopting model-independent techniques that utilize a calibrated $E_{\rm p}$--$E_{\rm iso}$ correlation for gamma-ray bursts (GRBs). To do so, in addition to real data points, we employ up to $1000$ simulated observational Hubble data (OHD) points. We then calibrate the $E_{\rm p}$--$E_{\rm iso}$ correlation by means of the well-consolidate B\'ezier polynomial technique, interpolating OHD up to the second order. Once GRB data have been calibrated, we consider two strategies of cosmographic expansions, i.e., first we take a direct Hubble rate expansion around $z_{tr}$, and second the expansion of the deceleration parameter around the same redshift, but with a different order. Employing type Ia supernovae, baryonic acoustic oscillations and GRB data sets, from Monte Carlo analyses we infer tight constraints on $z_{tr}$ and the jerk parameters at $z=z_{tr}$, namely $j_{tr}$. Our results are extremely compatible with previous outcomes and confirm the $\Lambda$CDM predictions, being slightly different in terms of the jerk parameter. In this respect, we conjecture which extensions of the concordance paradigm are possible and we compare our findings with expectations provided by generic dark energy models.   
\end{abstract}

\begin{keywords}
Gamma-ray Bursts:  general -- cosmology: Dark energy -- cosmology: Observations
\end{keywords}

\section{Introduction}
\label{intro}

The greatest challenge of modern cosmology is understanding the nature of dark matter and dark energy \citep{reviewDE1,Capozziello:2019cav} that constitute the $\sim25\%$ and $70\%$ of universe's content, respectively \citep{Planck2018}. A complete comprehension of the dark sector, namely both dark matter and dark energy, would shed light on the $\Lambda$CDM paradigm \citep{luomuc,2022NewAR..9501659P}. The concordance paradigm involves six free parameters, whereas at small redshifts can be easily approximated by a zero-spatial curvature, with a vacuum energy cosmological constant $\Lambda$, inferred from quantum field fluctuations \citep{weinberg}, and a uniform pressureless matter fluid, with negligible radiation, neutrino and relics abundances \citep{Planck2018}. 
While matter decelerates the expansion, $\Lambda$ acts as a bizarre repulsive gravity \citep{2016PDU....12...56B} that, instead, accelerates the universe. Consequently, there exists an \emph{onset of cosmic acceleration} at a given \emph{transition time} and/or \emph{transition redshift} \citep[see][and references therein]{2022MNRAS.509.5399C}. 

A fully clear scheme towards determining tight constraints on the transition time would suggest, in a model independent way, whether dark energy is under the form of a cosmological constant or it evolves in time\footnote{A cascade of alternative approaches has been proposed \citep{casc}, recently with a renewed aim to overcome cosmic tensions \citep[see, e.g.,][]{2020NatAs...4..196D,2021PhRvD.103d1301H}.
However, some of the models in \citet{casc} with barotropic index $\omega>-1$ are now at odds with the local Hubble constant $H_0$ values. The same occurs, at late times, for models where $\Lambda$ is replaced by a scalar field theories \citep{2021PhRvD.103h1305B,2022JCAP...04..004L}.} \citep[see, e.g.,][]{Chevallier2001,Linder2003,King2014}. The transition time, or alternatively the transition redshift $z_{tr}$, marks the era of dark energy domination over matter during which the universe enters a phase of accelerated expansion \citep{1973lsss.book.....H,2007PhRvD..76d1301M}. This quantity has  been recently reinterpreted as a genuine \emph{kinematic quantity}\footnote{We can distinguish the kinematics of the universe from its dynamics, once we assume metric expansions only, without directly involving Einstein's field equations \citep[see, e.g.,][]{Gruber:2013wua,Aviles:2014rma,2016IJGMM..1330002D,Capozziello:2017nbu}.}.

Thus, constraining $z_{tr}$ would prove in a model-independent manner how dark energy evolves and what consequences can be argued at the background without postulating the functional form of the Hubble parameter \citep[see, e.g.,][]{2016IJGMM..1330002D,Aviles:2012ay}. The strategy to write up $z_{tr}$ in terms of a given cosmological model has been proposed in \citet{2007PhRvD..76d1301M}, whereas characterising $z_{tr}$ without invoking an underlying cosmological model is proposed in \citet{2005GReGr..37.1541V}, \citet{2007CQGra..24.5985C,2008CQGra..25p5013C,2008PhRvD..78f3501C}, \citet{2012PhRvD..86l3516A} and \citet{Capozziello:2014zda,Capozziello:2015rda}. 

Another severe issue is that strategy of bounding $z_{tr}$ usually involves low redshift data, ignoring if the transition time is compatible with high-redshift data. In this respect, gamma-ray bursts (GRBs) represent plausible sources that, if used for cosmological purposes, may trace the universe expansion history in a more suitable way than using low and/or intermediate catalogs of data only \citep[see, e.g.,][]{2021ApJ...908..181M,2022MNRAS.512..439C,2022PASJ..tmp...83D,2022arXiv220809272J}.

In this work, we thus propose to constrain the transition time using the high-redshift GRB data, in order to check the compatibility of standard dark energy models at earlier times. To do so, we propose two fully model-independent strategies to reconstruct the Hubble rate $H(z)$ around $z=z_{tr}$. In particular, the first procedure makes use of a direct Hubble expansion, whereas the second expands the deceleration parameter $q$, that by definition vanishes at the transition time, namely $q_{tr}=q(z_{tr})=0$. The corresponding cosmological distances are modeled by calibrating the $E_{\rm p}$--$E_{\rm iso}$ GRB correlation. To do so, even in this case we formulate the model-independent calibration of the observed Hubble rate data, employing the widely-consolidate technique based on the B\'ezier polynomials. Hence, we calibrated our correlation adopting four hierarchies: first we utilized $32$ real data and then $100$, $500$ and $1000$ simulated ones. We therefore infer how error bars modify with the increase of the number of data points. Consequences on experimental analysis are also discussed, emphasizing the advantages of increasing the number of simulated data. 
In particular, we performed Markov chain Monte Carlo (MCMC) fits of our reconstructed Hubble expansions up to the jerk cosmographic parameter, inferred at $z=z_{tr}$, taking into account Pantheon type Ia supernovae (SNe Ia) and baryonic acoustic oscillarions (BAO) together with the most recent catalog of GRB data composed of $118$ long GRBs \citep{2021JCAP...09..042K}.

Finally, we debate about our fitting procedures and we show an overall compatibility of our results with respect to previous literature and the $\Lambda$CDM paradigm. Inside the intervals of validity for $z_{tr}$, we notice that evolving dark energy is not fully excluded. Consequences in physical cosmologies are therefore discussed. 

The paper is structured as follows. In Sec.~\ref{sezione2}, we describe the calibration based on the use of B\'ezier polynomials and the Amati relation. In the same section, we describe the method of simulating the observational data adopted as Hubble points. The corresponding analysis of simulated data is reported. In Sec.~\ref{sec:3}, we describe the two theoretical features related to our model-independent methods adopted to constrain the transition redshift. The experimental constraints of our simulations and the theoretical consequences of our results are described in Sec.~\ref{sec:4}. Finally, conclusions and perspectives are reported in Sec.~\ref{sezione5}.

\section{Model-independent calibration of the Amati relation through B\'ezier polynomials}\label{sezione2}

The most widely-used GRB correlation, named $E_{\rm p}$--$E_{\rm iso}$ or Amati correlation \citep{2002A&A...390...81A,AmatiDellaValle2013}, provides a relation between the peak and isotropic energies of GRBs. It reads
\begin{equation}
\label{Amatirel}
\log\left(\frac{E_{\rm p}}{{\rm keV}}\right)= a \left[\log\left(\frac{E_{\rm iso}}{{\rm erg}}\right)-52\right] + b\,,
\end{equation}
and it is defined by means of two parameters, namely the slope $a$ and the intercept $b$. The correlation is also characterized by an extra source of variability $\sigma$ \citep{Dago2005}.

In Eq.~\eqref{eiso}, $E_{\rm p}= E_{\rm p}^{\rm o}(1+z)$ is the observed peak energy $E_{\rm p}^{\rm o}$ of the $\gamma$-ray time-integrated $\nu F_\nu$ spectrum computed in the source rest-frame, whereas $E_{\rm iso}$ is the isotropic equivalent energy radiated in $\gamma$-rays, defined as \citep[see, e.g.,][for details]{2021Galax...9...77L}
\begin{equation}\label{eiso}
 E_{\rm iso}\equiv 4\pi d_{\rm L}^2 S_{\rm b}(1+z)^{-1}\,.
\end{equation}
The observed bolometric GRB fluence $S_{\rm b}$ is the integral of the $\nu F_\nu$ spectrum in the rest-frame $1-10^4$~keV energy band and the factor $(1+z)^{-1}$ transforms the GRB duration from the observer to the source cosmological rest-frame.
Here, we employ the most recent $E_{\rm p}$--$E_{\rm iso}$ catalog composed of $118$ long GRBs characterized by the smallest intrinsic dispersion \citep{2021JCAP...09..042K}.

We immediately notice that $E_{\rm iso}$ is determined once a background cosmology is imposed \emph{a priori} through the luminosity distance $d_{\rm L}$, leading to the well-known \emph{circularity problem}. 
To overcome this issue, we need to calibrate the $E_{\rm p}$--$E_{\rm iso}$ correlation by means of model-independent techniques involving SNe Ia \citep[see, e.g.,][]{2008ApJ...685..354L,Demianski2021} or average distance moduli got from different GRB correlations \citep[see, e.g.,][]{2015NewAR..67....1W}.
To this aim, we resort the well-established strategy of fitting OHD data by using a B\'ezier parametric curve \citep{2019MNRAS.486L..46A,LM2020,2021MNRAS.501.3515M}.
We utilize the updated sample of $32$ OHD \citep[see][and references therein]{2022arXiv220513247K} and interpolate it by employing a B\'ezier parametric curve of order $n$
\begin{equation}
\label{bezier1}
H_n(x)=\sum_{i=0}^{n} g_\alpha\alpha_i h_n^d(x)\quad,\quad h_n^i(x)\equiv n!\frac{x^i}{i!} \frac{\left(1-x\right)^{n-i}}{(n-i)!}\,,
\end{equation}
where $g_\alpha=100$~km/s/Mpc is a scaling factor and $\alpha_i$ are the coefficients of the linear combination of the polynomials $h_n^i(x)$, positive-defined for $0\leq x\equiv z/z_{\rm O}^{\rm m}\leq1$, with $z_{\rm O}^{\rm m}$ representing the maximum redshift of the OHD catalog. By construction, we identify $\alpha_0\equiv h_0=H_0/g_\alpha$.
\citet{LM2020} proved that only the function $H_2(z)$ (with $n=2$) is non-linear and monotonic growing with $z$ and can be extrapolated to $z>z_{\rm O}^{\rm m}$.
Therefore, supported by the finding $\Omega_k=0.001\pm0.002$ from \citet{Planck2018}, we can safely assume $\Omega_k=0$ and the luminosity distance becomes cosmology-independent\footnote{It is worth to mention that, according to recent claims on $\Omega_k$ representing at most the $2$\% of the total universe energy density \citep[see, e.g.,][and references therein]{2018ApJ...864...80O}, the circularity problem would not be completely healed, but would be only restricted to the value of $\Omega_k$ through $d_{\rm L}$.}
\begin{equation}
\label{dlHz2}
d_{\rm cal}(z)=c(1+z)\int_0^z\dfrac{dz'}{H_2(z')}\,.
\end{equation}
We are now in the position to use $d_{\rm cal}(z)$ to calibrate the isotropic energy $E_{\rm iso}^{\rm cal}$ for each GRB fulfilling the Amati relation
\begin{equation}
\label{Eisocal}
E_{\rm iso}^{\rm cal}(z)\equiv 4\pi d_{\rm cal}^2(z) S_{\rm b}(1+z)^{-1}\,,
\end{equation}
where the respective errors on $E_{\rm iso}^{\rm cal}$ depend upon the GRB systematics on the observables and the fitting procedure (see Sec.~\ref{sec:4}).

\begin{figure*}
\centering
\includegraphics[width=0.48\hsize,clip]{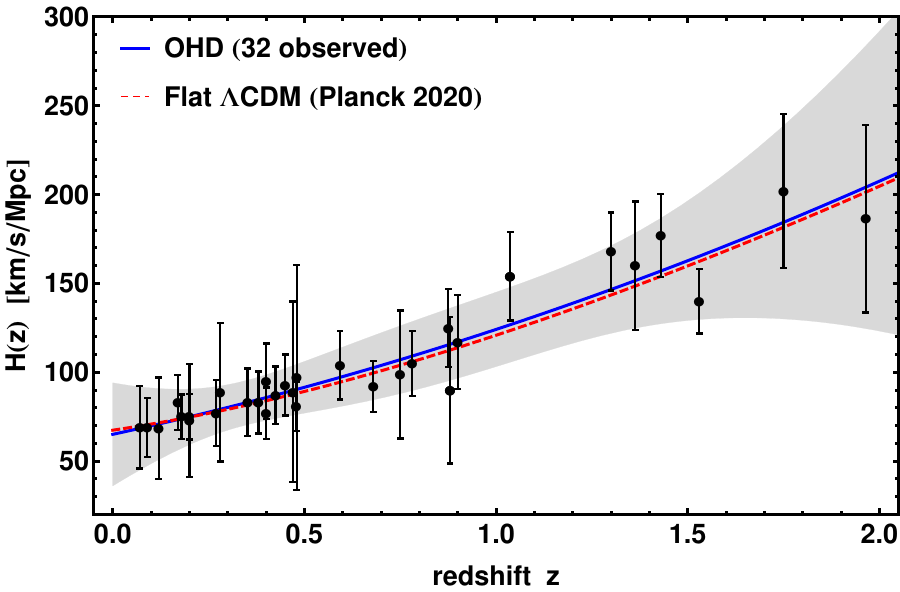}\hfill
\includegraphics[width=0.48\hsize,clip]{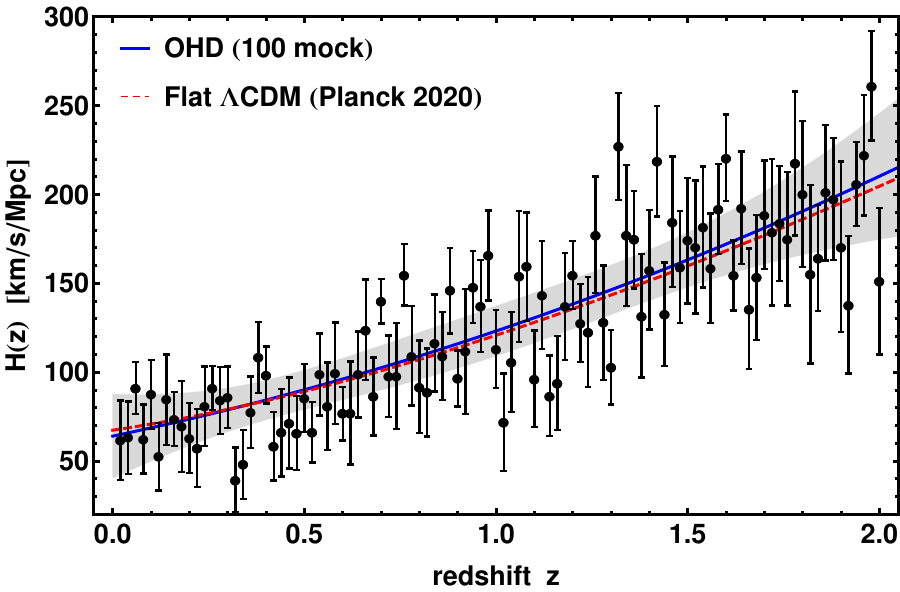}\hfill\\
\includegraphics[width=0.48\hsize,clip]{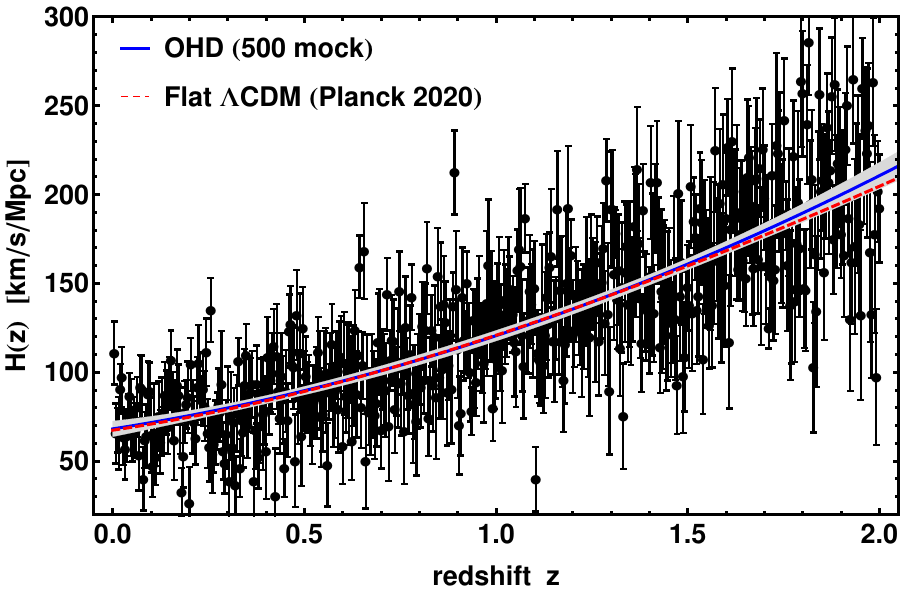}\hfill
\includegraphics[width=0.48\hsize,clip]{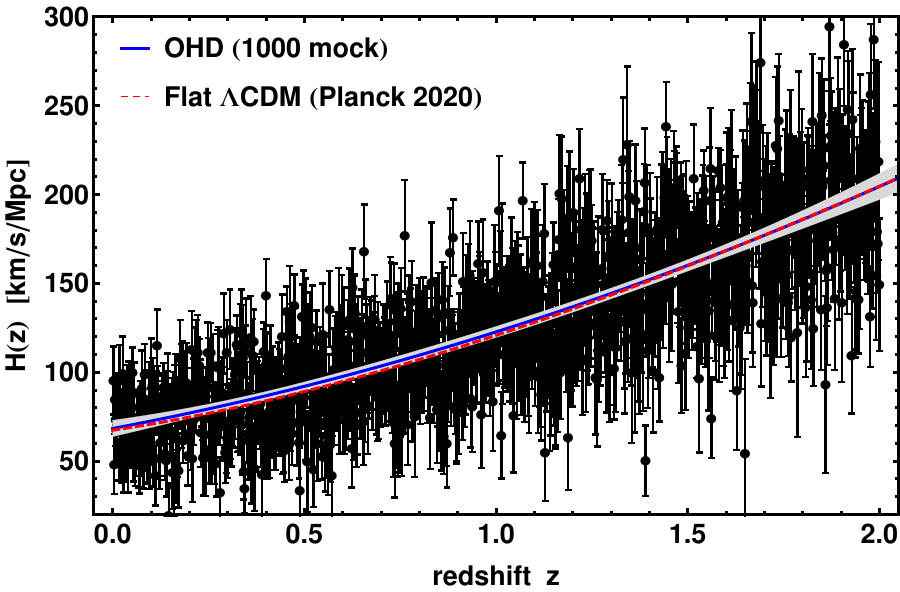}
\caption{Plot of the $H_2(z)$ B\'ezier curves (blue thick lines) with the $3$--$\sigma$ confidence bands (gray shaded area) compared with the $\Lambda$CDM paradigm (dashed red curves). The data employed are: $32$ real OHD measurements (top left) and $100$ (top right), $500$ (bottom left), and $1000$ mock OHD points (bottom right).}
\label{fig:Bez}
\end{figure*}
\begin{table*}
\centering
\setlength{\tabcolsep}{.9em}
\renewcommand{\arraystretch}{1.3}
\begin{tabular}{lccccccccc}
\hline\hline
OHD data                        & \vline &
$\alpha_0$                  & 
$\alpha_1$                  & 
$\alpha_2$                  & 
$z_{\rm tr}$                &\vline &
$\delta\alpha_0^{\rm min}$--$\delta\alpha_0^{\rm max}$ & 
$\delta\alpha_1^{\rm min}$--$\delta\alpha_1^{\rm max}$ & 
$\delta\alpha_2^{\rm min}$--$\delta\alpha_2^{\rm max}$ \\
\hline
$32$ real                   & \vline &
$0.651^{+0.091}_{-0.092}$   & 
$1.111^{+0.219}_{-0.223}$   & 
$2.042^{+0.248}_{-0.246}$   & 
$0.618^{+0.078}_{-0.133}$   &\vline &
$0.140$--$0.142$   & 
$0.200$--$0.200$   & 
$0.121$--$0.120$   \\
$100$ mock                   & \vline &
$0.641^{+0.060}_{-0.062}$   & 
$1.092^{+0.112}_{-0.113}$   & 
$2.103^{+0.089}_{-0.094}$   &
$0.534^{+0.011}_{-0.046}$   &\vline &
$0.094$--$0.096$   & 
$0.102$--$0.103$   & 
$0.043$--$0.045$   \\
$500$ mock                  & \vline &
$0.681^{+0.024}_{-0.024}$   & 
$1.018^{+0.051}_{-0.047}$   & 
$2.106^{+0.040}_{-0.042}$   &
$0.684^{+0.002}_{-0.002}$   & \vline &
$0.035$--$0.036$   & 
$0.050$--$0.047$   & 
$0.019$--$0.020$   \\
$1000$ mock                 & \vline &
$0.684^{+0.017}_{-0.018}$   & 
$1.089^{+0.035}_{-0.033}$   & 
$2.045^{+0.026}_{-0.029}$   &
$0.738^{+0.001}_{-0.001}$   & \vline &
$0.024$--$0.026$   & 
$0.032$--$0.030$   & 
$0.013$--$0.014$   \\
\hline
\end{tabular}
\caption{Best-fit B\'ezier coefficients $\alpha_i$ and relative maximum and minimum error bars of the function $H_2(z)$. The inferred $z_{\rm tr}$ is also shown.}
\label{tab:Bez} 
\end{table*}

\begin{figure*}
\centering
\includegraphics[width=0.33\hsize,clip]{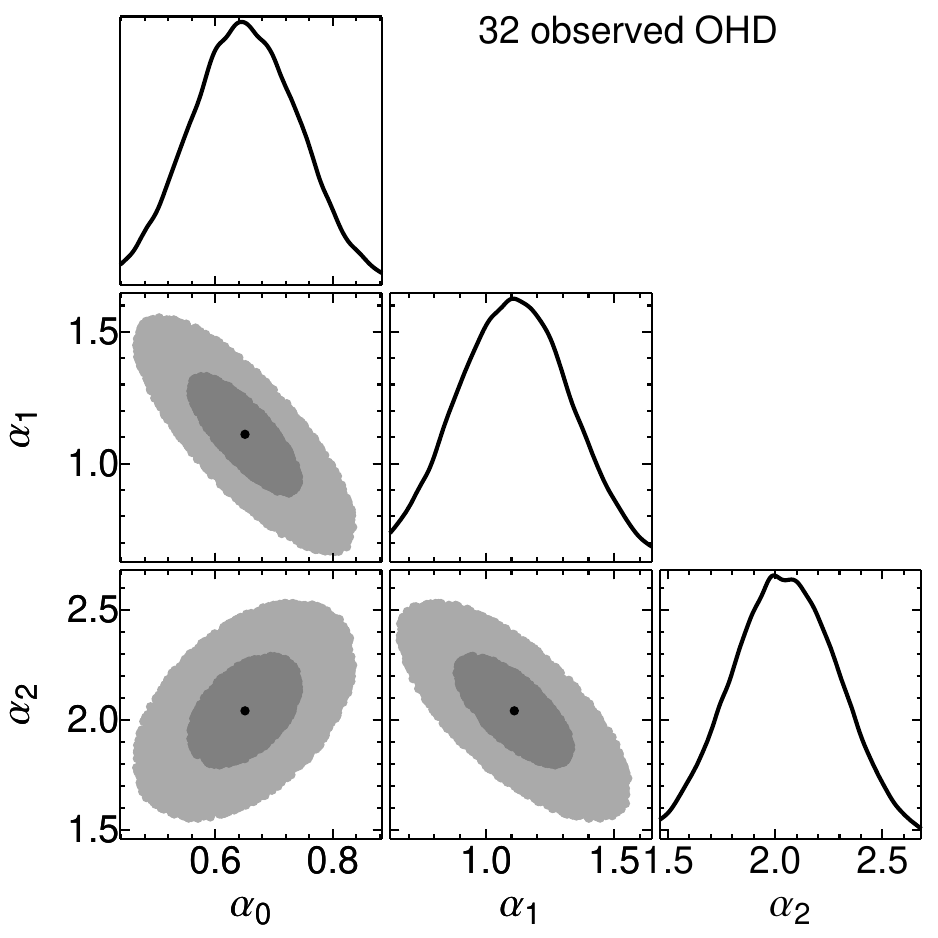}
\hfil
\includegraphics[width=0.33\hsize,clip]{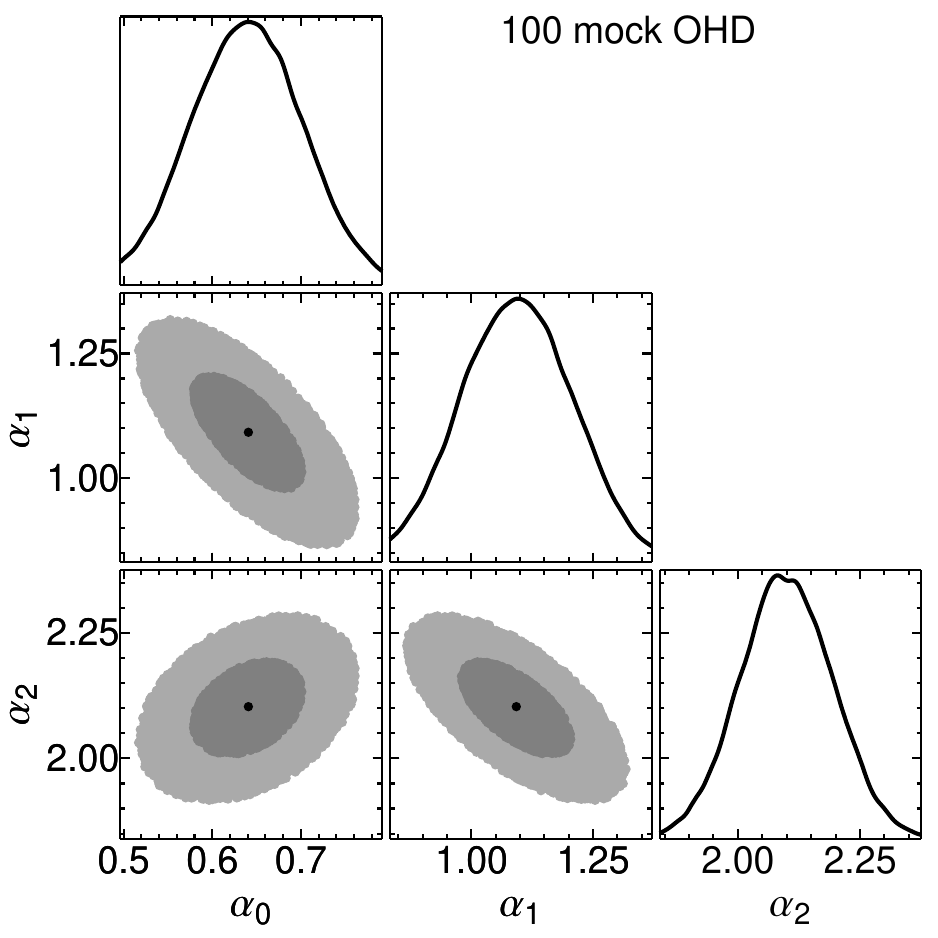}\hfil\\
\includegraphics[width=0.33\hsize,clip]{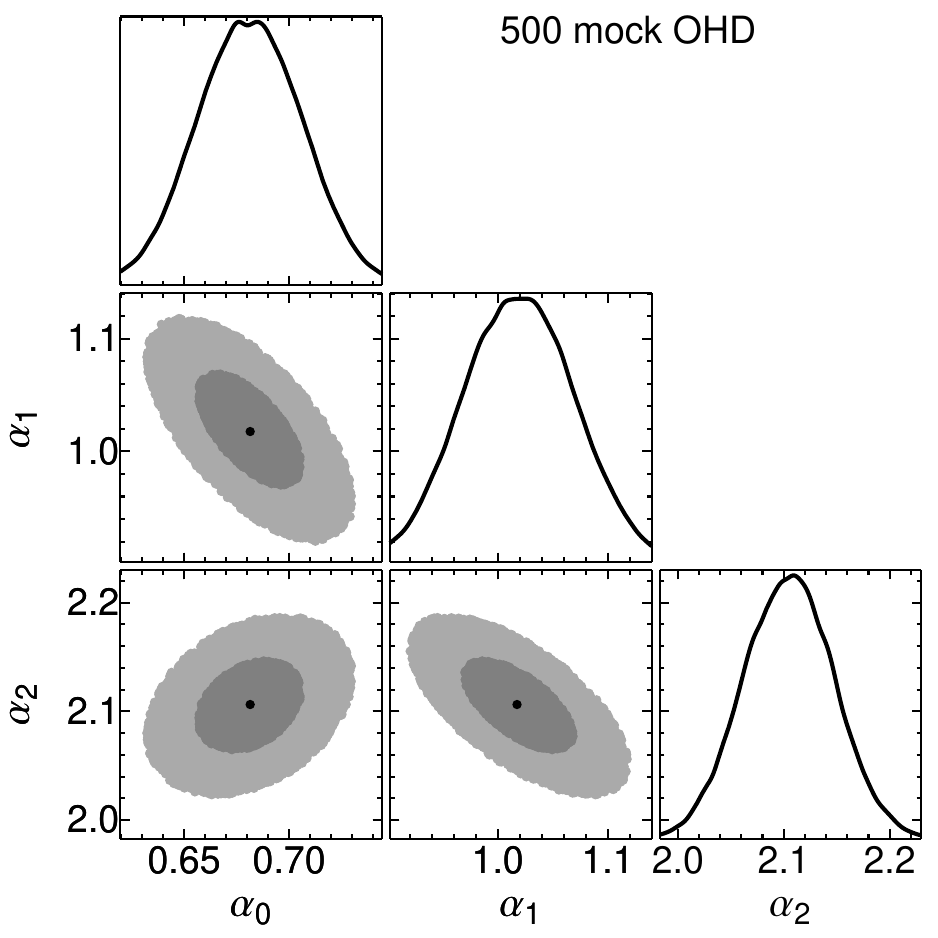}
\hfil
\includegraphics[width=0.33\hsize,clip]{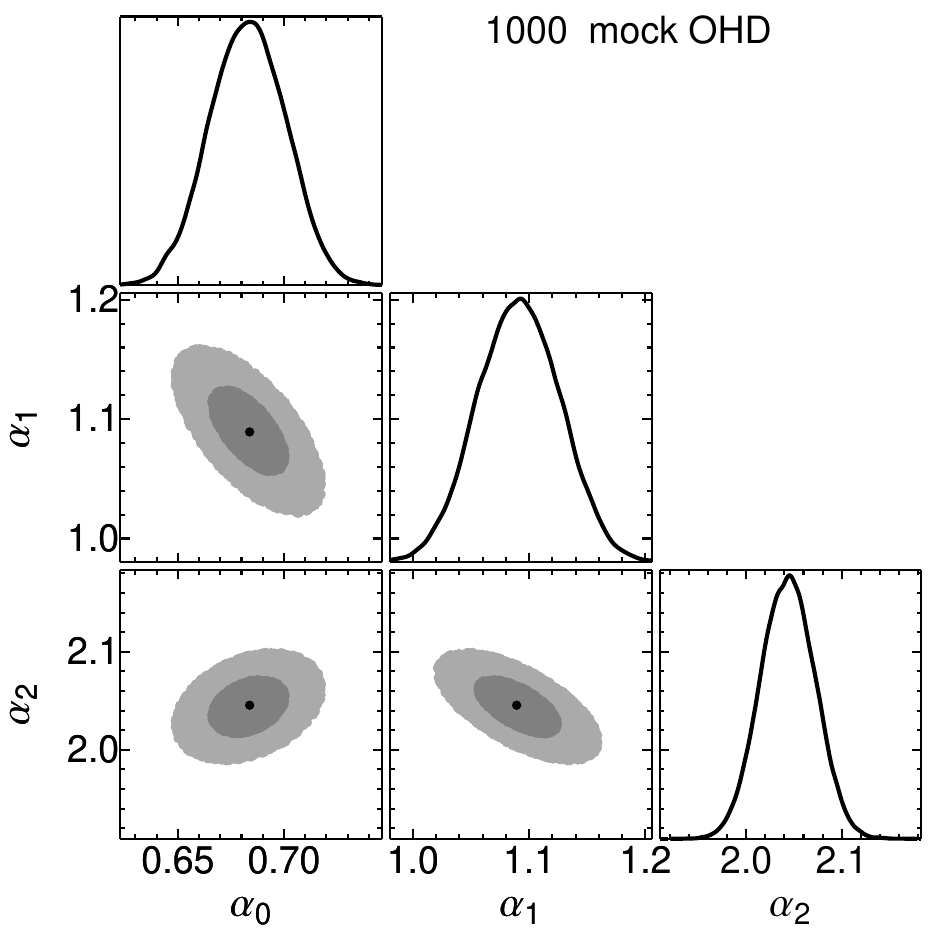}
\caption{Contour plots of the best-fit B\'ezier coefficients $\alpha_i$ of $H_2(z)$ for real and simulated OHD. Darker (lighter) areas mark $1$--$\sigma$ ($2$--$\sigma$) confidence regions.}
\label{fig:Bez_cont}
\end{figure*}

\subsection{Method of simulating Hubble rate data}

In this new era of precision cosmology, we expect the improvement in both the quantity and quality of data. 
Currently, the main limitations for OHD measurements are the large errors and the absence of dedicated surveys.
Hence, to explore the effect of a large number of data points, we need to simulate the Hubble data.
\citet{2022LRR....25....6M} evaluated the impact of a moderate number of Hubble points from the available spectroscopic surveys at low redshifts, assuming statistical errors of $1\%$ and systematic errors as suggested by \citet{2020ApJ...898...82M}, and from future spectroscopic surveys at high redshifts, by simulating measurements with statistical error of $5\%$. Here, we keep the same statistical and systematic errors of the current spectroscopic surveys and only increase the size of the catalog by simulating OHD, as suggested by \citet{2011ApJ...730...74M}, in the redshift range $0 < z < 2$. 

Before doing so, we perform a careful evaluation of the systematic uncertainties of OHD measurements.
In \citet{2020ApJ...898...82M,2022LRR....25....6M}, the total OHD covariance matrix is given by the sum of the contributions of statistical and systematic errors
\begin{equation}
{\rm Cov}_{ij} = {\rm Cov}_{ij}^{\rm stat} + {\rm Cov}_{ij}^{\rm syst} \,.
\end{equation}
The systematic part of the covariance matrix is decomposed
into\begin{equation}
\label{eq:Covtot}
{\rm Cov}_{ij}^{\rm syst} = {\rm Cov}_{ij}^{\rm young} + {\rm Cov}_{ij}^{\rm met} + {\rm Cov}_{ij}^{\rm mod}\,,
\end{equation}
where the sources of error depend on, respectively, the possible residual star formation due to a \textit{young} subdominant component underlying in the selected galaxy sample, the estimate of the stellar \textit{metallicity} of the population, and the considered \textit{model}.
The last contribution, in turn, can be further decomposed in
\begin{equation}
\label{eq:Covmod}
{\rm Cov}_{ij}^{\rm mod} = {\rm Cov}_{ij}^{\rm SPS} + {\rm Cov}_{ij}^{\rm st. lib.} + {\rm Cov}_{ij}^{\rm IMF} + {\rm Cov}_{ij}^{\rm SFH}\,,
\end{equation}
which depends on the \textit{stellar population synthesis} (SPS) model used to calibrate the measurement that involves stellar physics models, together with the adopted \textit{stellar library}, the \textit{initial mass function} (IMF), and the \textit{star formation history} (SFH).

${\rm Cov}_{ij}^{\rm stat}$, ${\rm Cov}_{ij}^{\rm met}$, ${\rm Cov}_{ij}^{\rm SFH}$ and ${\rm Cov}_{ij}^{\rm young}$ are diagonal matrixes, since they are  uncorrelated for objects at different redshifts \citep{2022LRR....25....6M}.
The terms of ${\rm Cov}_{ij}^{\rm stat}$ can be taken from \citet{2022arXiv220513247K} or \citep{2022LRR....25....6M}.
The contributions of ${\rm Cov}_{ij}^{\rm SFH}$ and ${\rm Cov}_{ij}^{\rm young}$ has been shown to be negligible \citep{2020ApJ...898...82M}, whereas the contributions of ${\rm Cov}_{ij}^{\rm SPS}$, ${\rm Cov}_{ij}^{\rm st. lib.}$, and ${\rm Cov}_{ij}^{\rm IMF}$ can be found in the form of mean percentage offsets $\hat\eta^k(z)$ as functions of the redshift from the second to the fourth columns\footnote{We assume that OHD measurements at $z>1.475$ have the same mean offsets attached to $z=1.475$.} of Table~$3$ in \citep{2020ApJ...898...82M}.
The average effect due to metallicity (conservatively assumed to a $5\%$ accuracy) is $4.16\%$ \citep[see Table~$4$ in][]{2020ApJ...898...82M}.

The full systematic covariance matrix of the most relevant contributions due to the model is given by 
\begin{equation}
{\rm Cov}_{ij}^{\rm mod} = \sum_k\hat\eta^k(z_i)\times H(z_i) \times \hat\eta^k(z_j)\times H(z_j)
\end{equation}
where $k$ indicates IMF, stellar library, and SPS, see Eq.~\eqref{eq:Covmod}.

Finally, the total covariance in Eq.~\eqref{eq:Covtot} can be used to attach the errors to OHD measurements and to build up simulated larger OHD catalogs, following the steps briefly outlined below.
\begin{itemize}
    \item[1.] First, we choose the flat $\Lambda$CDM model as a fiducial model with matter density parameter $\Omega_{\rm m0} = 0.3153\pm0.0073$ and Hubble constant $H_0 = (67.36\pm0.54)$~km/s/Mpc \citep{Planck2018}.
    \item[2.] Next, we define the deviation $\Delta H$ between the simulated and the fiducial Hubble data points, \emph{i.e.},
    $$ H_{\rm sim}(z) = H_{\rm fid} + \Delta H\,.$$
    \item[3.] Then, we plot the uncertainties of the observed $H(z)$ data points with $z$. After removing $7$ outliers, we interpolate the errors with two lines: $\sigma_+(z) = 16.577 z + 18.440$ for the upper errors and $\sigma_-(z) = 7.402 z + 2.675$ for the lower errors. Hence, to simulate the Hubble data points, we define $\sigma_{\rm m}$ as the mean of $\sigma_+$ and $\sigma_-$. 
    \item[4.] By assuming a Gaussian distribution $\mathcal G$, the random uncertainty $\sigma_{\rm r}(z)$ in simulating the $H(z)$ points can be estimated as $\mathcal G( \sigma_{\rm m}, \sigma_{\rm v})$, where the variance $\sigma_{\rm v}=(\sigma_+ -\sigma_-)/4$ ensures that $\sigma_{\rm r}(z)$ lies between $\sigma_+ $ and $\sigma_-$ at $95.4\%$ probability.
    \item[5.] Finally, the deviation $\Delta H$ is now obtained by assuming the Gaussian distribution $\mathcal G[0,\sigma_{\rm r}(z)]$.
    \item[6.] The above steps generate the simulated Hubble data points $H_{\rm sim}$ with the associated uncertainty given by $\sigma_{\rm r}(z)$ at uniform redshift bins in the range $0<z<2$.
\end{itemize}

\subsection{Direct calibration}

We are now in the position to estimate the coefficients $\alpha_i$ ($0\leq i\leq2$) of the B\'ezier curve $H_2(z)$.
Assuming Gaussian distributed errors, the OHD the log-likelihood function reads as
\begin{equation}
\label{loglikeOHD}
    \ln \mathcal{L}_{\rm O} = -\frac{1}{2}\sum_{k=1}^{N_{\rm O}}\left\{\left[\dfrac{H_k-H_2(z_k)}{\sigma_{H,k}}\right]^2 + \ln(2\pi\,\sigma_{H,k}^2)\right\}\,,
\end{equation}
where $N_{\rm O}$ is the size of the OHD catalog with values $H_k$ and attached errors $\sigma_{H,k}$. We here consider various OHD catalogs: $32$ real measurements, and $100$, $500$ and $1000$ mock data.

The best-fit curves approximating the various OHD catalogs are portrayed in Fig.~\ref{fig:Bez}, where a comparison with the predictions of the $\Lambda$CDM paradigm \citep{Planck2018} is also shown.
The best-fit coefficients $\alpha_i$ used in Fig.~\ref{fig:Bez} are summarized in Table~\ref{tab:Bez} and displayed in the contour plots of Fig.~\ref{fig:Bez_cont}.
In Table~\ref{tab:Bez} are also listed the values of the transition redshift $z_{\rm tr}$ inferred from each (real and mock) OHD catalog. These values have been found from the corresponding functions $H_2(z)$ imposing the condition on the deceleration parameter $q(z_{\rm tr})\equiv0$, namely
\begin{equation}
\label{qzH2}
q(z_{\rm tr}) = \left\{\frac{(1+z)^2}{H_2(z)}\frac{d}{dz}\left[\frac{H_2(z)}{1+z}\right]\right\}_{z=z_{\rm tr}}\equiv0
\end{equation}

\begin{figure}
\centering
\includegraphics[width=0.99\hsize,clip]{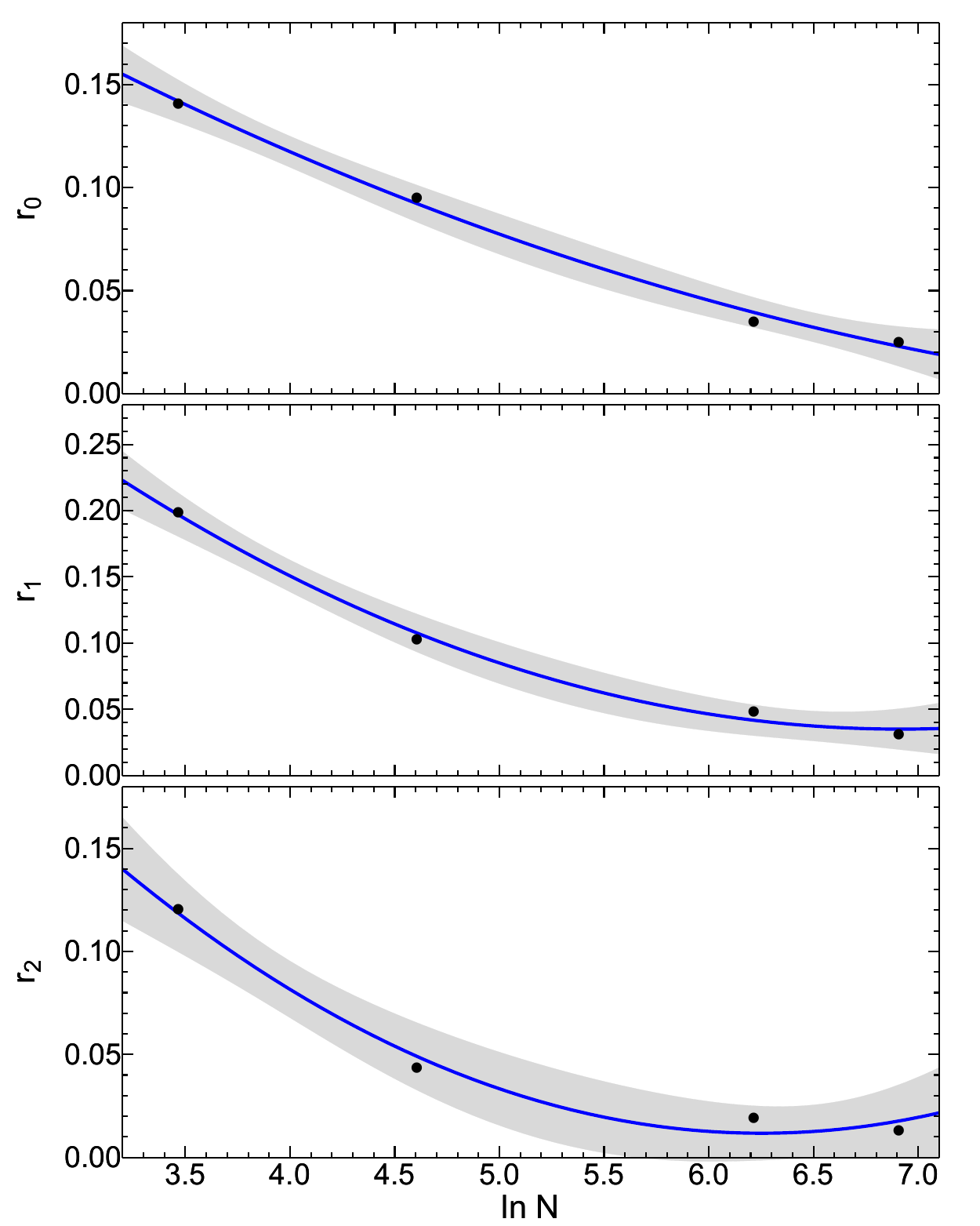}
\caption{Plots of the mean value between minimum ad maximum error bars $r_i\equiv(\delta\alpha_i^{\rm min}+\delta\alpha_i^{\rm max})/2$ as a function of the natural logarithm of the number points, namely $N=\{32,\,100,\,500,\,1000\}$ as reported in Table \ref{tab:Bez}. Light gray areas mark the $1$--$\sigma$ confidence bands.}
\label{fig:alphadecrease}
\end{figure}

Once the error bars are evaluated, we computed the error evolution as data points increase. 
In particular, in Fig.~\ref{fig:alphadecrease} we draw plots of the relative errors $\delta\alpha_i=\sigma_{\alpha,i}/\alpha_i$ decrease, with $0\leq i\leq2$. There each point corresponds to the mean value of the relative error defined by $r_i\equiv(\delta\alpha_i^{\rm min}+\delta\alpha_i^{\rm max})/2$ at the corresponding value of the natural logarithm of the number points, namely $N=\{32,\,100,\,500,\,1000\}$ as shown in Table \ref{tab:Bez}.

The functional behavior of the error decrease is written as 
\begin{equation}
r_i(\ln N) = \mathcal A_{i} + \mathcal B_{i} \ln N + \mathcal C_{i} (\ln N)^2\,,
\end{equation}
where the free coefficients have been evaluated by the \texttt{FindFit} command in \texttt{Wolfram Mathematica} and read
\begin{subequations}
\begin{align}
\mathcal A_{i}&\simeq\{0.356,\,0.0686,\,0.550\},\\ 
\mathcal B_{i}&\simeq\{-0.075,\,-0.188,\,-0.172\},\\ 
\mathcal C_{i}&\simeq\{-0.004,\,0.013,\,0.014\}\,,    
\end{align}
\end{subequations}
where the first values of $\mathcal A_i$, $\mathcal B_i$ and $\mathcal C_i$ refer to $r_0$ and so forth.
All the aforementioned cases show that the decreases are slightly approximated by second order polynomials, though the second order is subdominant with respect to the first one. The steepest decrease is associated to $\delta\alpha_1$, whereas the largest uncertainty is attached to $\delta\alpha_2$. The theoretical treatment shows that error bars tend to smaller values as our simulated data are included. The large discrepancy between expectations got from the real catalog and the simulated ones is a proper indication that it is not suitable to match together both simulated and real data. For this reason, we split the overall analysis by considering the values reported in Table ~\ref{tab:Bez}. 

\section{Model-independent Hubble rate and transition redshifts}\label{sec:3}

The transition redshift can be constrained adopting two model-independent procedures. The first strategy is constructed by expanding the Hubble rate around $z=z_{tr}$, while the second approach is analogous but expands directly the deceleration parameter. The two methods are below reported \citep[see][for details]{2022MNRAS.509.5399C}.

\begin{itemize}
    \item[--] \emph{{\bf Direct Hubble Expansion (DHE) method}}. Expanding the Hubble rate, we immediately get
\begin{equation}
\label{primometodoHespanso}
H=H_{tr}+H_{tr}^\prime(z-z_{tr})+\frac{1}{2}H_{tr}^{''}(z-z_{tr})^2+\mathcal O(z-z_{tr})^3\,,
\end{equation}
where the additional normalizing request on 
$H(z=0)\equiv H_0$ relates the cosmographic parameters among them. In particular, since%
\begin{equation}
\label{Hpunto}
\dot{H} = -H^2 (1 + q)\quad,\quad
\ddot{H}  = H^3 (j + 3q + 2)\,,
\end{equation}
and $a\equiv(1+z)^{-1}$, we obtain
\begin{equation}
H^{''}_{tr}=\frac{H_{tr}j_{tr}}{(1+z_{tr})^2}\quad,\quad
H^\prime_{tr}=\frac{H_{tr}}{1+z_{tr}}\,,
\end{equation}
with 
\begin{equation}
    H_{tr} = \frac{2 H_0 (1 + z_{tr})^2}{ 2 + z_{tr} (2 + j_{tr} z_{tr})}\,,
\end{equation}
providing as final result the normalized Hubble rate that reads
\begin{equation}\label{2primometodoHespanso2}
\mathcal E_{\rm DHE}(z)=1+\frac{2 z (1 + z_{tr} - j_{tr} z_{tr})+j_{tr} z^2}{2 +
 z_{tr} (2 + j_{tr} z_{tr})}\,.
\end{equation}
\item[--] \emph{{\bf Direct Deceleration Parameter Expansion (DDPE) method}}. This second appoach directly expands the deceleration parameter up to a given order around  $z_{tr}$. We here limit to first Taylor order of expansion to have
\begin{equation}\label{eq:exp q}
q\simeq q_{tr}+q_{tr}^\prime(z-z_{tr})+\mathcal O(z-z_{tr})^2\,,
\end{equation}
where we recall $q_{tr}=0$, noticing to be able to get $q_0$, by simply taking $z=0$ inside Eq.~\eqref{eq:exp q}.

In particular, the deceleration parameter is negative definite to guarantee current acceleration, when $z\leq z_{tr}$ since 
the transition occurs when $q$ passes from a positive to a negative value. It is thus needful to get the Hubble rate and  so we substitute Eq.~\eqref{eq:exp q} inside the second Friedmann equation\footnote{Remarkably, if one  expands $H$ around $z=0$  and plugs Eqs. \eqref{eq:exp q} with $z=0$ would lead to worse results in computation.}, fixing $H(z=0)=H_0$, to obtain 
\begin{equation}
\label{Hmetodo2ordine2e3}
\mathcal E_{\rm DDPE}(z)=\exp\left(\frac{j_{tr} z}{1 + z_{tr}}\right) (1 + z)^{1 - j_{tr}}\,.
\end{equation}
\end{itemize}

\begin{table*}
\centering
\setlength{\tabcolsep}{.4em}
\renewcommand{\arraystretch}{1.6}
\begin{tabular}{lccccccc}
\hline\hline
Calibration                 &
$a$                         & 
$b$                         & 
$\sigma$                    &
$h_0$                       &
$z_{\rm tr}$                &
$j_{\rm tr}$                \\    
\hline
\multicolumn{7}{c}{Model DHE}\\		
$32$ real                       &
$0.737^{+0.031 (+0.077)}_{-0.033 (-0.066)}$   & 
$1.784^{+0.048 (+0.101)}_{-0.049 (-0.109)}$   & 
$0.289^{+0.026 (+0.050)}_{-0.011 (-0.031)}$   &
$0.692^{+0.008 (+0.017)}_{-0.009 (-0.018)}$   &
$0.702^{+0.094 (+0.198)}_{-0.044 (-0.105)}$   &
$1.096^{+0.109 (+0.285)}_{-0.175 (-0.338)}$  \\
$1000$ mock                     &
$0.738^{+0.035(+0.074)}_{-0.035(-0.067)}$       &
$1.790^{+0.043(+0.092)}_{-0.050(-0.111)}$       &
$0.294^{+0.018(+0.046)}_{-0.016(-0.035)}$       &
$0.692^{+0.007(+0.016)}_{-0.009(-0.018)}$       &
$0.726^{+0.065(+0.171)}_{-0.072(-0.127)}$       &
$1.057^{+0.147(+0.332)}_{-0.133(-0.293)}$       \\
\hline
\multicolumn{7}{c}{Model DDPE}\\
$32$ real                       &
$0.738^{+0.033(+0.077)}_{-0.036(-0.066)}$   &
$1.784^{+0.049(+0.095)}_{-0.057(-0.112)}$   &
$0.292^{+0.023(+0.049)}_{-0.016(-0.033)}$   &
$0.688^{+0.011(+0.020)}_{-0.005(-0.014)}$   &
$0.806^{+0.083(+0.187)}_{-0.051(-0.105)}$   &
$1.057^{+0.133(+0.257)}_{-0.105(-0.246)}$   \\
$1000$ mock                     &
$0.743^{+0.027(+0.073)}_{-0.042(-0.072)}$   &
$1.784^{+0.047(+0.095)}_{-0.048(-0.107)}$   &
$0.300^{+0.015(+0.042)}_{-0.022(-0.040)}$   &
$0.691^{+0.009(+0.017)}_{-0.008(-0.017)}$   &
$0.820^{+0.067(+0.177)}_{-0.067(-0.116)}$   &
$1.070^{+0.120(+0.242)}_{-0.126(-0.260)}$   \\
\hline
\end{tabular}
\caption{Best-fit coefficients of the calibrated $E_{\rm p}$--$E_{\rm iso}$ correlation, calibrated via the $32$ real and the $1000$ mock OHD measurements, and best-fit parameters of the DHE and DDPE methods.}
\label{tab:fits} 
\end{table*}

\begin{figure*}
\centering
\includegraphics[width=0.498\hsize,clip]{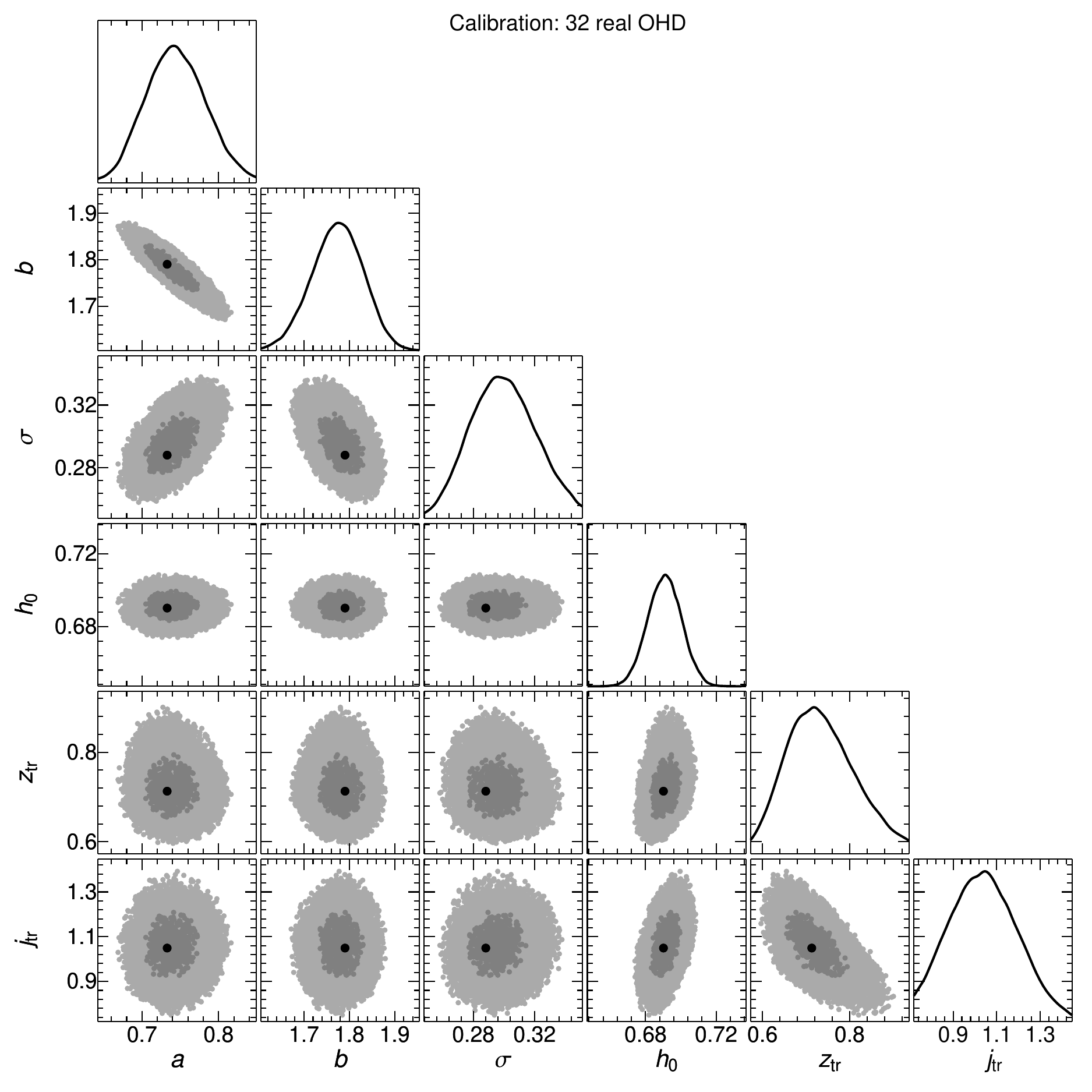}
\includegraphics[width=0.498\hsize,clip]{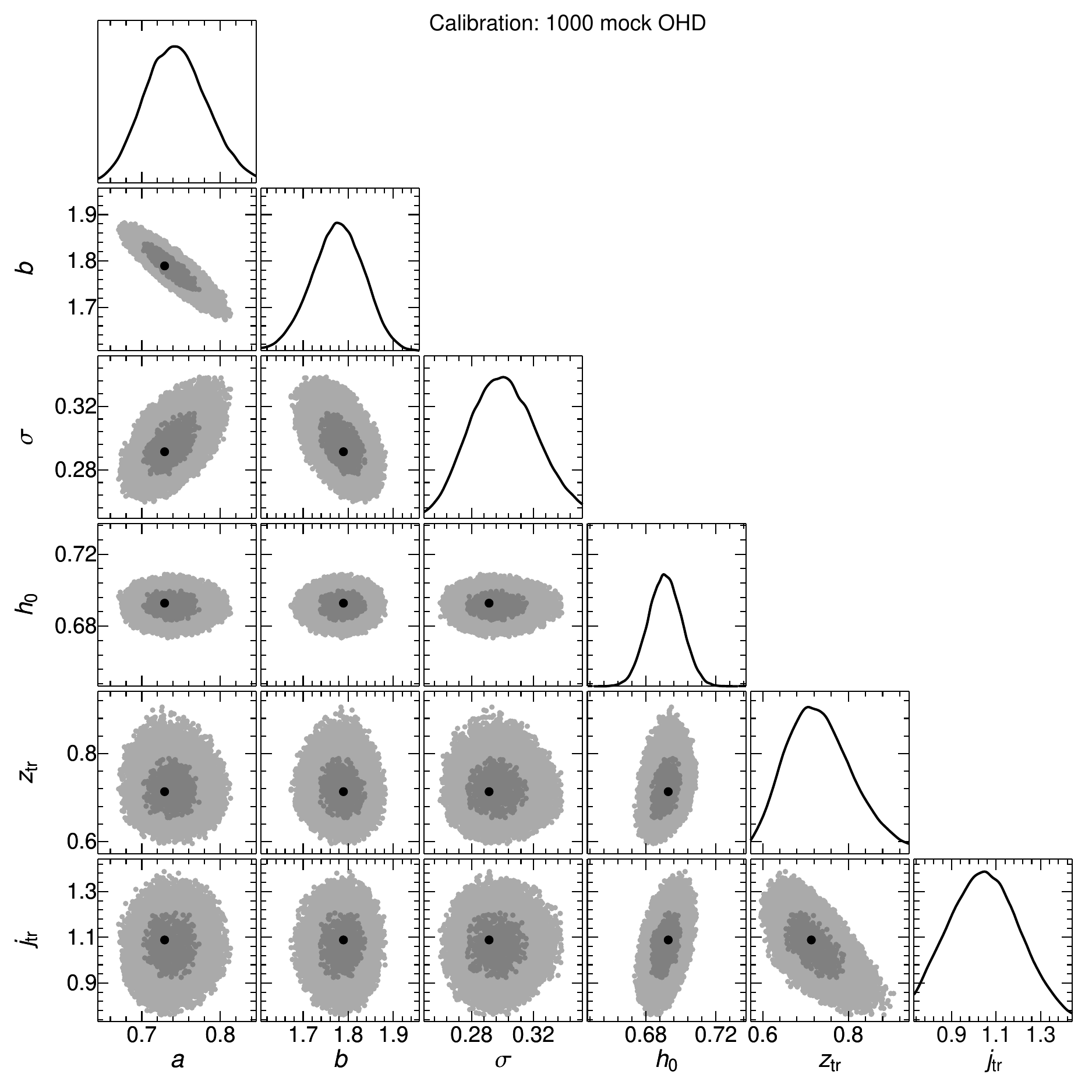}
\caption{Contour plots of the best-fit coefficients of the $E_{\rm p}$--$E_{\rm iso}$ correlation, calibrated via the $32$ real and the $1000$ mock OHD measurements, and the best-fit parameters of the DHE method. Darker (lighter) areas mark the $1$--$\sigma$ ($2$--$\sigma$) confidence regions.}
\label{fig:Bez_cont2}
\end{figure*}

\begin{figure*}
\centering
\includegraphics[width=0.498\hsize,clip]{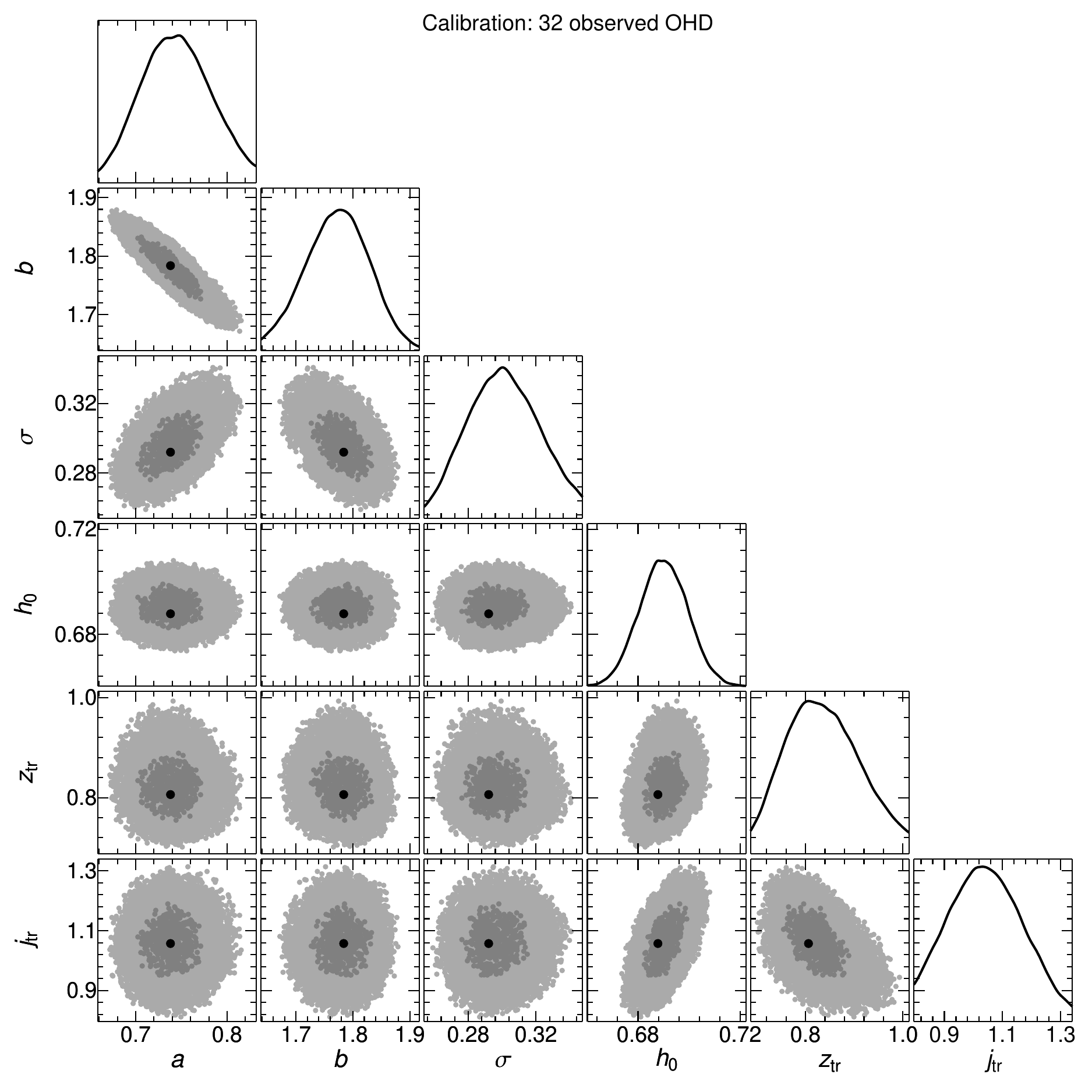}
\includegraphics[width=0.498\hsize,clip]{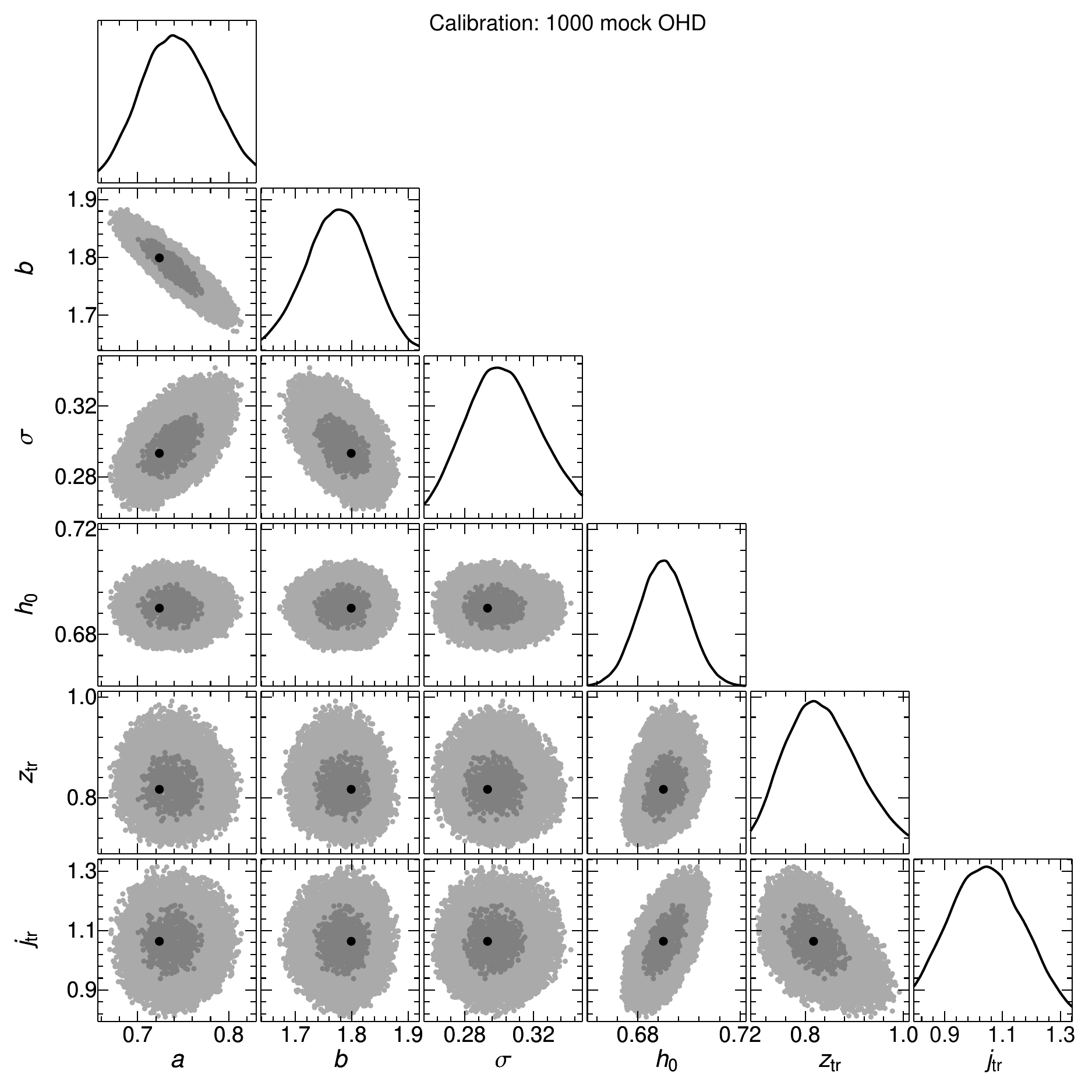}
\caption{The same as Fig.~\ref{fig:Bez_cont2} but for the DDPE method.}
\label{fig:Bez_cont3}
\end{figure*}

\subsection{Impact of Taylor expansions}

The above expansions are essentially \emph{cosmographic series} of the Hubble rate and deceleration parameter \citep{Luongo:2015zgq}. Arguably, one can think that increasing the expansion order of those series would improve the quality of the corresponding fits. The situation is, however, different due to the \emph{convergence problem} \citep{Capozziello:2020ctn}. The latter likely represents the largest limitation of cosmographic expansions. Indeed, truncating series at a given expansion order increases the issue related to the fact that quite all cosmological data sets exceed the bound at which our above expansions have been computed. 

All Taylor series exhibit mathematical divergences as the redshift increases. Consequently, as $z\rightarrow\infty$, the so-constructed  polynomials in Eqs. \eqref{primometodoHespanso} and \eqref{eq:exp q} tend to dramatically deviate from a suitable approximation. This is the case of GRBs, i.e., the larger redshift domains influences the finite truncation, inducing severe systematic errors in computing the final output of our findings. 

The corresponding bad convergence, that may affect numerical results, has been carefully taken into account in the above expansions. Indeed, the DHE can arrive up to the second order in deriving $H$, while the DDPE to an apparent lower order in expanding $q$ that, however, corresponds to the same of $H$, namely up to the cosmographic jerk  \citep{2017PDU....17...25A}. So, following \citet{Luongo:2011zz}, and employing the fact that up to jerk order, cosmographic expansions appear more predictive than further orders, we assumed the aforementioned expansions in Eqs. \eqref{primometodoHespanso} and \eqref{eq:exp q}. 

Further approximations, namely higher Taylor expansions, would cause higher error propagation that would severely plague our findings. Even though this caveat appears to limit the kind of approximations that one can perform, strategies toward its resolutions have been also discussed. Among all, re-parameterizing  the redshift variable, alternatives to Taylor expansions, etc. \citep[see e.g.,][]{2018JCAP...05..008C,Luongo:2015zgq,Gruber:2013wua}. However every possibility consists in stringent (non-divergent) alternative series that are not treated in detail in this paper and will be investigated in future developments.

\section{Experimental constraints}\label{sec:4}

Here, we perform MCMC analyses to find out the set of parameters entering the total log-likelihood function
\begin{equation}
 \ln{\mathcal{L}} = \ln{\mathcal{L}_{\rm G}} + \ln{\mathcal{L}_{\rm S}} + \ln{\mathcal{L}_{\rm B}}\,.
\end{equation}
Below, we describe each contribution and for each of them we assume Gaussian distributed errors.

\subsection{GRB log-likelihood}
To calibrate the GRB data through the OHD interpolation and to get cosmological bounds from it, we follow a hierarchical Bayesian regression \citep{2021A&A...647A..72K} or nested log-likelihood approach \citep{LM2020} in which the GRB log-likelihood $\ln{\mathcal{L}_{\rm G}}$ is determined from the modeling of the combination of two sub-samples: 
\begin{itemize}
    \item[i)]  a \textit{calibrator sample} of GRBs encompassing OHD observations up to $z^{\rm m}_{\rm O}$, used to determine the correlation coefficients, and 
    \item[ii)] a \textit{cosmological sample} of the whole GRB data set, used to estimate the free parameters of the model.
\end{itemize}
It is worth to mention that, though employed for calibration purposes, the GRBs of the calibrator sample are not detracted from whole cosmological one. This has been done for two important reasons. On the one hand, if the calibrator sample is excluded from the cosmological one, then there would be a drastic cut in redshift ($z<2$) of the GRB catalog, where essentially the constraints on $h_0$ and $z_{tr}$ come from. On the other hand, as we are going to see below, the OHD distance-corrected calibrator sample employs $E_{\rm p}$--$E_{\rm iso}^{\rm cal}$ pairs to estimate the correlation parameters, whereas the cosmological sample utilizes yet different $E_{\rm p}$--$S_{\rm b}$ pairs to extract the cosmological bounds.

Within the Bayesian approach, the posterior of the model parameters $\theta=\{a,b,\sigma,h_0,z_{\rm tr},j_{\rm tr}\}$ is given by
\begin{equation}
\label{eq:bayes}
 P(\theta | \xi) \propto P(\xi| \theta) P(\theta),
\end{equation}
where $\xi=\{E_{\rm p},E_{\rm iso}^{\rm cal},S_{\rm b}\}$ are GRB observables. Each GRB provides its own set of poorly constrained model parameters $\theta_i$. Hence, to use the whole GRB sample in a consistent way, it is necessary to propagate hierarchically the model assumptions and priors for each source. With this prescription, we can marginalize over all $\theta_i$ assuming their distribution to be Gaussian with a mean given by $\theta$. This leads to the definition of the likelihood probability distribution $P(\xi|\theta)$ which is the combination the calibrator sample, with $N_{\rm cal}$ GRBs, and the cosmological sample, with $N_{\rm cos}$ GRBs,
\begin{equation}
    P(\xi| \theta) = \prod_{i=1}^{N_{\rm cal}} P[\xi_i , d_{\rm cal}(z_i) | \theta]  \prod_{j=1}^{N_{\rm cos}} P(\xi_j, z_j | \theta)\,.
\end{equation}
Finally, the term $P(\theta)$ in Eq.~\eqref{eq:bayes} gives the (uniform) priors on the model parameters.

Therefore, the total GRB log-likelihood function is given by
\begin{equation}
\label{a0}
    \ln \mathcal{L}_{\rm G} = \ln \mathcal{L}_{\rm G}^{\rm cal} + \ln \mathcal{L}_{\rm G}^{\rm cos}\,.
\end{equation}
The calibration log-likelihood is given by
\begin{equation}
\label{a1}
\ln \mathcal{L}_{\rm G}^{\rm cal} = -\frac{1}{2}\sum_{i=1}^{N_{\rm cal}}\left\{\left[\dfrac{Y_i-Y(z_i)}{\sigma_{ Y,i}}\right]^2 + \ln(2\pi\,\sigma_{Y,i}^2)\right\}\,,\\
\end{equation}
where $N_{\rm cal}=65$ and
\begin{subequations}
\begin{align}
Y_{\rm i} \equiv&\, \log E_{{\rm p},i}\,,\\
\label{a2}
Y(z_i)\equiv&\, a \left[\log E_{\rm iso}^{\rm cal}(z_i)-52 \right] + b\,,\\
\sigma_{Y,i}^2 \equiv&\, \sigma_{\log E_{{\rm p},i}}^2 + a^2\sigma_{\log[E_{\rm iso}^{\rm cal}(z_i)]}^2+\sigma^2\,.
\end{align}
\end{subequations}
The cosmological log-likelihood is given by
\begin{equation}
\label{a4}
\ln \mathcal{L}_{\rm G}^{\rm cos} = -\frac{1}{2}\sum_{j=1}^{N_{\rm cos}}\left\{\left[\dfrac{\mu_j-\mu_{\rm th}(z_j)}{\sigma_{\mu,j}}\right]^2 + \ln (2 \pi \,\sigma_{\mu,j}^2)\right\}\,, 
\end{equation}
where we have $N_{\rm cos}=118$ and
\begin{subequations}
\begin{align}
\label{a5}
\mu_j\equiv& \,\frac{5}{2 a}\left[\log E_{{\rm p},j} - a\log\left(\frac{4\pi S_{{\rm b},j}}{1+z_j}\right) - b\right]\,,\\
\sigma_{\mu,j}^2 \equiv& \,\frac{25}{4 a^2}\left(\sigma_{\log E_{{\rm p},j}}^2 + a^2 \sigma_{\log S_{{\rm b},j}}^2+\sigma^2\right)\,,
\end{align}
\end{subequations}
where the theoretical distance modulus is given by
\begin{equation}
\label{muz}
\mu_{\rm th}(z)= 25+5\log\left[\frac{d_{\rm L}(z)}{{\rm Mpc}}\right]\,.
\end{equation}
Assuming $\Omega_k=0$, the luminosity distance of the model reads
\begin{equation}
\label{dlEz}
d_{\rm L}(z)=\frac{c}{H_0}(1+z)\int_0^z\dfrac{dz'}{\mathcal E(z')}\,,
\end{equation}
with $\mathcal E(z)$ given by DHE or DDPE models.

\subsection{SN log-likelihood}
The Pantheon data set is the most updated SN Ia sample composed of $1048$ sources \citep{2018ApJ...859..101S}.
In \citet{2018ApJ...853..126R}, such data points are prompted under the form of $\mathcal E(z)^{-1}$ catalog at $N_{\rm S}=6$ redshifts, chosen to best represent the whole SN Ia sample and obtained by assuming a flat universe prior.
The SN log-likelihood function is given by
\begin{align}
\nonumber
\ln \mathcal{L}_{\rm S} = & -\frac{1}{2}\sum_{k=1}^{N_{\rm S}} \left[\mathcal E_k^{-1} - \mathcal E(z_k)^{-1} \right]^{\rm T} \mathbf{C}_{\rm SN}^{-1}
\left[\mathcal E_k^{-1} - \mathcal E(z_k)^{-1} \right]\\
\label{loglikeSN}
& -\frac{1}{2}\sum_{k=1}^{N_{\rm S}} \ln \left(2 \pi |\det\mathbf{C}_{\rm SN}| \right)\,, 
\end{align}
where $\mathcal E_k^{-1}$ are the measured from SNe Ia and $\mathbf{C}_{\rm SN}$ is the covariance matrix obtained from the correlation matrix in \citet{2018ApJ...853..126R}.
The six $\mathcal E_k(z)$ measurements ensure great computational time compression and accurately reproduce the cosmological constraints of the whole ($>1000$) SN Ia data set \citep{2018ApJ...853..126R}.

\begin{table*}
\centering
\setlength{\tabcolsep}{.5em}
\renewcommand{\arraystretch}{1.5}
\begin{tabular}{cccccccc}
\hline\hline
$z$ bin                     &
$a$                         & 
$b$                         & 
$\sigma$                    &
$h_0$                       &
$z_{\rm tr}$                &
$j_{\rm tr}$                \\    
\hline
\multicolumn{7}{c}{Model DHE}\\
\hline
\multicolumn{7}{c}{Calibration: $32$ real}\\
$0.0$--$2.0$                       &
$0.689^{+0.061 (+0.129)}_{-0.038 (-0.085)}$   & 
$1.864^{+0.050 (+0.121)}_{-0.091 (-0.187)}$   & 
$0.277^{+0.035 (+0.078)}_{-0.021 (-0.045)}$   &
$0.694^{+0.007 (+0.016)}_{-0.011 (-0.019)}$   &
$0.700^{+0.083 (+0.174)}_{-0.063 (-0.119)}$   &
$1.082^{+0.176 (+0.331)}_{-0.140 (-0.295)}$  \\
$2.0$--$3.0$                       &
$0.730^{+0.079 (+0.144)}_{-0.032 (-0.077)}$   & 
$1.804^{+0.054 (+0.122)}_{-0.090 (-0.176)}$   & 
$0.319^{+0.044 (+0.086)}_{-0.022 (-0.047)}$   &
$0.690^{+0.010 (+0.020)}_{-0.006 (-0.015)}$   &
$0.692^{+0.076 (+0.186)}_{-0.048 (-0.104)}$   &
$1.146^{+0.111 (+0.275)}_{-0.207 (-0.364)}$  \\
$3.0$--$8.2$                       &
$0.743^{+0.073 (+0.139)}_{-0.043 (-0.100)}$   & 
$1.830^{+0.059 (+0.141)}_{-0.104 (-0.210)}$   & 
$0.317^{+0.033 (+0.081)}_{-0.031 (-0.061)}$   &
$0.693^{+0.007 (+0.016)}_{-0.009 (-0.018)}$   &
$0.706^{+0.082 (+0.175)}_{-0.047 (-0.113)}$   &
$1.102^{+0.112 (+0.275)}_{-0.169 (-0.330)}$  \\
\multicolumn{7}{c}{Calibration: $1000$ mock}\\	
$0.0$--$2.0$                     &
$0.757^{+0.037 (+0.105)}_{-0.069 (-0.117)}$   & 
$1.768^{+0.088 (+0.160)}_{-0.048 (-0.131)}$   & 
$0.334^{+0.026 (+0.073)}_{-0.034 (-0.060)}$   &
$0.691^{+0.009 (+0.018)}_{-0.008 (-0.016)}$   &
$0.707^{+0.066 (+0.170)}_{-0.069 (-0.122)}$   &
$1.079^{+0.173 (+0.351)}_{-0.131 (-0.292)}$  \\
$2.0$--$3.0$   &
$0.697^{+0.056 (+0.120)}_{-0.054 (-0.099)}$   & 
$1.845^{+0.081 (+0.143)}_{-0.064 (-0.154)}$   & 
$0.280^{+0.034 (+0.073)}_{-0.022 (-0.049)}$   &
$0.690^{+0.010 (+0.018)}_{-0.007 (-0.015)}$   &
$0.685^{+0.096 (+0.192)}_{-0.046 (-0.101)}$   &
$1.094^{+0.146 (+0.317)}_{-0.146 (-0.303)}$  \\
$3.0$--$8.2$   &
$0.749^{+0.049 (+0.123)}_{-0.068 (-0.123)}$   & 
$1.814^{+0.091 (+0.168)}_{-0.068 (-0.181)}$   & 
$0.312^{+0.037 (+0.086)}_{-0.030 (-0.055)}$   &
$0.691^{+0.009 (+0.019)}_{-0.007 (-0.016)}$   &
$0.691^{+0.094 (+0.203)}_{-0.040 (-0.098)}$   &
$1.098^{+0.129 (+0.284)}_{-0.161 (-0.320)}$  \\
\hline
\multicolumn{7}{c}{Model DDPE}\\
\hline
\multicolumn{7}{c}{Calibration: $32$ real}\\
$0.0$--$2.0$                       &
$0.748^{+0.044(+0.117)}_{-0.059(-0.107)}$   &
$1.770^{+0.077(+0.156)}_{-0.062(-0.144)}$   &
$0.332^{+0.029(+0.072)}_{-0.033(-0.060)}$   &
$0.689^{+0.010(+0.020)}_{-0.006(-0.015)}$   &
$0.784^{+0.075(+0.173)}_{-0.044(-0.100)}$   &
$1.086^{+0.134(+0.274)}_{-0.095(-0.239)}$   \\
$2.0$--$3.0$                       &
$0.716^{+0.038(+0.106)}_{-0.064(-0.119)}$   &
$1.829^{+0.081(+0.153)}_{-0.062(-0.147)}$   &
$0.286^{+0.027(+0.068)}_{-0.029(-0.054)}$   &
$0.692^{+0.008(+0.017)}_{-0.010(-0.017)}$   &
$0.790^{+0.078(+0.171)}_{-0.044(-0.101)}$   &
$1.131^{+0.082(+0.221)}_{-0.162(-0.283)}$   \\
$3.0$--$8.2$                       &
$0.734^{+0.066(+0.142)}_{-0.045(-0.097)}$   &
$1.830^{+0.068(+0.151)}_{-0.096(-0.198)}$   &
$0.311^{+0.041(+0.087)}_{-0.024(-0.050)}$   &
$0.691^{+0.008(+0.018)}_{-0.008(-0.017)}$   &
$0.806^{+0.075(+0.177)}_{-0.052(-0.108)}$   &
$1.097^{+0.091(+0.235)}_{-0.131(-0.271)}$   \\
\multicolumn{7}{c}{Calibration: $1000$ mock}\\
$0.0$--$2.0$                     &
$0.744^{+0.051(+0.110)}_{-0.060(-0.106)}$   &
$1.797^{+0.059(+0.128)}_{-0.072(-0.165)}$   &
$0.330^{+0.034(+0.075)}_{-0.032(-0.057)}$   &
$0.691^{+0.009(+0.018)}_{-0.007(-0.017)}$   &
$0.801^{+0.064(+0.161)}_{-0.059(-0.116)}$   &
$1.074^{+0.143(+0.281)}_{-0.085(-0.237)}$   \\
$2.0$--$3.0$                     &
$0.686^{+0.065(+0.130)}_{-0.044(-0.093)}$   &
$1.853^{+0.068(+0.135)}_{-0.073(-0.167)}$   &
$0.279^{+0.035(+0.074)}_{-0.021(-0.047)}$   &
$0.691^{+0.009(+0.018)}_{-0.007(-0.017)}$   &
$0.795^{+0.069(+0.165)}_{-0.059(-0.112)}$   &
$1.127^{+0.098(+0.225)}_{-0.147(-0.275)}$   \\
$3.0$--$8.2$                     &
$0.742^{+0.052(+0.128)}_{-0.055(-0.114)}$   &
$1.828^{+0.075(+0.160)}_{-0.087(-0.184)}$   &
$0.324^{+0.032(+0.077)}_{-0.033(-0.064)}$   &
$0.693^{+0.005(+0.015)}_{-0.010(-0.019)}$   &
$0.808^{+0.074(+0.168)}_{-0.059(-0.110)}$   &
$1.105^{+0.089(+0.234)}_{-0.144(-0.274)}$   \\
\hline
\end{tabular}
\caption{Best-fit coefficients of the calibrated $E_{\rm p}$--$E_{\rm iso}$ correlation, calibrated via the $32$ real and the $1000$ mock OHD measurements, and best-fit parameters of the DHE and DDPE methods for three redshift bins.}
\label{tab:zbins} 
\end{table*}

\subsection{BAO log-likelihood} 
We select $N_{\rm B}=8$ uncorrelated BAO angular distance measurements \citep[see, e.g.,][]{LM2020} being not explicitly dependent upon $\Omega_{\rm m0}$, \emph{i.e.},
\begin{equation}
\label{eq:DV}
\Delta(z) \equiv r_{\rm s} \left[\frac{H(z)}{cz}\right]^\frac{1}{3}\left[\frac{\left(1+z\right)}{d_{\rm L}(z)}\right]^\frac{2}{3}\,,
\end{equation}
where $r_{\rm s}$ is the comoving sound horizon at the baryon drag redshift $z_\text{d}$, calibrated through CMB data for a given cosmological model\footnote{In this sense BAO are slightly model-dependent.}. Hereafter we consider $r_{\rm s}=(147.21\pm0.48)$~Mpc \citep{Planck2018}.
The corresponding log-likelihood is given by
\begin{equation}
\label{loglikebao}
\ln \mathcal{L}_{\rm B} = -\frac{1}{2}\sum_{k=1}^{N_{\rm B}}\left\{\left[\frac{\Delta_k - \Delta(z_k)}{\sigma_{\Delta,k}}\right]^2 + \ln (2 \pi \,\sigma_{\Delta,k}^2)\right\}\,. 
\end{equation}

\subsection{Numerical results}
In view of the results displayed in Fig.~\ref{fig:Bez} and summarized in Table~\ref{tab:Bez}, it is clear that the $1000$ mock OHD data well reproduce the results obtained from the $32$ real measurements, of course with greater accuracy.
Therefore, for comparison, we decided to calibrate GRBs using the real $32$ and the mock $1000$ OHD.

We performed MCMC fits, working out the Metropolis-Hastings algorithm and using the DHE and the DDPE methods outlined in Sec.~\ref{sec:3}, by means of a modified free available \texttt{Wolfram Mathematica} code \citep{codice}. 
The results summarized in Table ~\ref{tab:fits} are displayed in the contour plots of Figs.~\ref{fig:Bez_cont2}--\ref{fig:Bez_cont3}.
In  Table ~\ref{tab:zbins} we also checked and verified the stability of the correlation and cosmological parameters
in different redshift bins containing roughly the same number of GRBs.

MCMC joint calibration and cosmological fits lead to best-fit parameters (and associated errors) that are insensitive to the OHD data sets employed for calibrating $E_{\rm p}$--$E_{\rm iso}$ data.
This seems quite unexpected, because the $1000$ mock OHD data provide a more accurate reconstruction of $H(z)$, compared to that from the $32$ real measurements.
However, this is reasonable if one keeps in mind that, among the cosmological probes considered in the MCMC fits, GRBs are the ones with the largest errors, therefore,  their weight in determining the best-fit results is certainly minor with respect to SNe Ia and BAO data. 
This situation is more accentuated by the calibration procedure, because GRB errors become even larger due to OHD uncertainties that propagate via $d_{\rm cal}(z)$.

Nonetheless, the insensitivity to OHD measurements is only apparent because the values of $h_0$ we have found from our MCMC fits (see Table~\ref{tab:fits}) are consistent with those got from OHD (see $\alpha_0$ in Table~\ref{tab:Bez}), which are incorporated in the GRB calibrated data\footnote{The intercept parameter $b$ of the $E_{\rm p}$--$E_{\rm iso}$ relation degenerates with $h_0$. However, once calibrated via OHD, this degeneracy no longer holds and the joint fit with BAO data enables to get constraints on $h_0$.}, and BAO \citep{2019JCAP...10..044C}.

Finally, we notice that the estimates of $z_{\rm tr}$ obtained with the DHE method (see Table~\ref{tab:fits}) are in agreement with the corresponding ones obtained directly from the $H_2(z)$ functions of the real $32$ and the $1000$ mock OHD measurements (see Table~\ref{tab:Bez}). Contrary, this is not verified for the DDPE method, which provides larger and inconsistent values of $z_{\rm tr}$.

\subsection{Theoretical discussion of the results}
\label{sec:de_evol}

Our experimental results show a suitable compatibility toward the parameters $a$, $b$ and $\sigma$, concerning the Amati calibration. The model-independent calibration, based on B\'ezier polynomials, provides viable constraints that are quite compatible with previous findings got from the literature \citep[see, e.g.,][]{2019MNRAS.486L..46A,LM2020,2021JCAP...09..042K}. Moreover, the bounds show smaller error bars, consequently, also reducing the relative errors over the free coefficients that are needful for calibrating GRBs. The values of Hubble rates today are stable and provide the same mean outcomes for both DHE and DDPE methods. Errors are comparable between the two approaches, being surprisingly small even at the $2$--$\sigma$ confidence levels. The corresponding Hubble constant values are compatible, within the error bars, with Planck measurements \citep{Planck2018}, but far from Riess expectations \citep{2019ApJ...876...85R}, leaving unsolved the $H_0$ tension. The transition redshift appears larger and closer to $z\simeq1$ for the DDPE method, but highly compatible with the standard model predictions given by the standard cosmological model. Error bars are large enough at $2\sigma$ confidence levels to recover the $\Lambda$CDM expectations even considering the DDPE method. Surprisingly, the values of $j_{tr}$ got from our experimental analyses are extremely close to $j=1$ provided by the standard cosmological model. This shows that any possible departures from the concordance paradigm is forced to be extremely small, leading to a very weakly evolving dark energy term even if one involves high redshift data sets, such as GRB data. 

To better show the latter prediction that we argue from our analyses, we can consider the following generic Hubble rate
\begin{equation}\label{hz}
H(z)=H_0\sqrt{\Omega_{m0}(1+z)^{3}+\Omega_{DE}F(z)}\,,
\end{equation}
where the generic function $F(z)$ fulfills the following requirements%
\begin{equation}
\left\{
        \begin{array}{lll}
         F(z)\rightarrow 1 \quad&,\quad & \hbox{$z=0$}\\
        \Omega_{DE}\equiv1-\Omega_{m0}\quad&,\quad & \hbox{$\forall z$}\\
        \Omega_{DE}F(z)\gtrsim\Omega_{m0}(1+z)^3\quad&,\quad & \hbox{$z\rightarrow 0$}
        \end{array}
\right.
\end{equation}
that guarantee $H=H_0$ as $z=0$ and dark energy dominating over matter at late-times. 
Considering Eq.~\eqref{hz}, immediately we get
\begin{subequations}
\begin{align}
\label{qudef}
q(z)&=-1+\frac{(1+z)\left[3\Omega_{m0}(1+z)^{2}+\Omega_{DE}F^\prime(z)\right]}{2\left[\Omega_{m0}(1+z)^{3}+\Omega_{DE}F(z)\right]}\,,\\
\label{jeidef}
j(z)&=1+\frac{\Omega_{DE}(1+z)\left[-2 F^\prime(z)+(1+z)F^{\prime\prime}(z)\right]}{2\left[\Omega_{m0}(1+z)^{3}+\Omega_{DE}F(z)\right]}\,.
\end{align}
\end{subequations}
At the transition redshift Eqs.~\eqref{qudef}--\eqref{jeidef} lead to  
\begin{subequations}
\begin{align}
\label{Fpdef}
F^\prime_{\rm tr}&=-\frac{\Omega_{m0}(1+z_{\rm tr})^2}{1-\Omega_{m0}} + \frac{2 F_{\rm tr}}{1+z_{\rm tr}}\,,\\
\label{Fppdef}
F^{\prime\prime}_{\rm tr}&=-\frac{2\Omega_{m0}(2-j_{\rm tr})(1+z_{\rm tr})}{1-\Omega_{m0}} + \frac{2 (1+j_{\rm tr}) F_{\rm tr}}{(1+z_{\rm tr})^2}\,,
\end{align}
\end{subequations}
where we imposed $q_{\rm tr}=0$ and set $F_{\rm tr}=F(z_{\rm tr})$, $F^\prime_{\rm tr}=F^\prime(z_{\rm tr})$ and $F^{\prime\prime}_{\rm tr}=F^{\prime\prime}(z_{\rm tr})$.

To fully constrain the dark energy evolution, we need a further equation for $F_{\rm tr}$.
Using the B\'ezier interpolation of the Hubble rate and Eq.~\eqref{hz} and imposing that $H_2(z_{\rm tr})\equiv H(z_{\rm tr})$, we get 
\begin{subequations}
\begin{align}
\label{Fdef}
F_{\rm tr}&=\frac{H_2^2(z_{\rm tr})-H_0^2\Omega_{m0}(1+z_{\rm tr})^3}{H_0^2(1-\Omega_{m0})}\,,\\
\label{Fpdef2}
F^\prime_{\rm tr}&=\frac{2H_2^2(z_{\rm tr})-3H_0^2\Omega_{m0}(1+z_{\rm tr})^3}{H_0^2(1-\Omega_{m0})(1+z_{\rm tr})}\,,\\
\label{Fppdef2}
F^{\prime\prime}_{\rm tr}&=\frac{2H_2^2(z_{\rm tr})(1+j_{\rm tr})-6H_0^2\Omega_{m0}(1+z_{\rm tr})^3}{H_0^2(1-\Omega_{m0})(1+z_{\rm tr})^2}\,.
\end{align}
\end{subequations}
Taking the constraints in Tab.~\ref{tab:fits}, got for both DHE and DDPE methods, and the value of $\Omega_{m0}$ from \citet{Planck2018}, we obtain the constraints on $F_{\rm tr}$, $F^\prime_{\rm tr}$ and $F^{\prime\prime}_{\rm tr}$ summarized in Table.~\ref{tab:Ftr}.
\begin{table}
\centering
\setlength{\tabcolsep}{.5em}
\renewcommand{\arraystretch}{1.6}
\begin{tabular}{llccccc}
\hline\hline
Model                       & 
Calibration                 &
$F_{\rm tr}$                       &
$F^\prime_{\rm tr}$                &
$F^{\prime\prime}_{\rm tr}$\\                
\hline
DHE                         & 
real $32$                   &
$0.978^{+0.563}_{-0.419}$  & 
$-0.186^{+0.670}_{-0.560}$   & 
$-0.003^{+1.016}_{-0.768}$  \\
                            & 
mock $1000$                 &
$1.000^{+0.284}_{-0.314}$   & 
$-0.214^{+0.379}_{-0.426}$  & 
$-0.118^{+0.640}_{-0.676}$   \\
\hline
DDPE                        & 
real $32$                   &
$1.011^{+0.608}_{-0.495}$  &
$-0.383^{+0.711}_{-0.580}$   & 
$-0.294^{+0.944}_{-0.728}$   &  \\
                            & 
mock $1000$                 &
$0.987^{+0.329}_{-0.324}$   & 
$-0.441^{+0.424}_{-0.412}$  & 
$-0.326^{+0.617}_{-0.617}$  \\
\hline
\end{tabular}
\caption{Constraints on the dark energy evolution through $F_{\rm tr}$, $F^\prime_{\rm tr}$ and $F^{\prime\prime}_{\rm tr}$, for DHE and DDPE methods and calibrations based on the $32$ real and the $1000$ mock OHD measurements.}
\label{tab:Ftr} 
\end{table}

It appears evident from Table~\ref{tab:Ftr} that the value of $F_{tr}$ is compatible with the constraint $F_{tr}=1$ even at $1\sigma$ confidence level. However, the first derivative evaluated at the transition time is not compatible with zero, whereas the second derivative does it. This implies that our expectations agree with those of a slightly evolving dark energy term that however departs from a genuine $\Lambda$CDM scenario, albeit recovering it at small redshifts. This finding may be due to the need of a further constraint that we imposed by means of B\'ezier polynomials. It is likely that imposing a tighter bound over $F(z)$ would imply tighter constraints probably more compatible with the standard cosmological model.

\begin{figure*}
\centering
\includegraphics[width=0.48\hsize,clip]{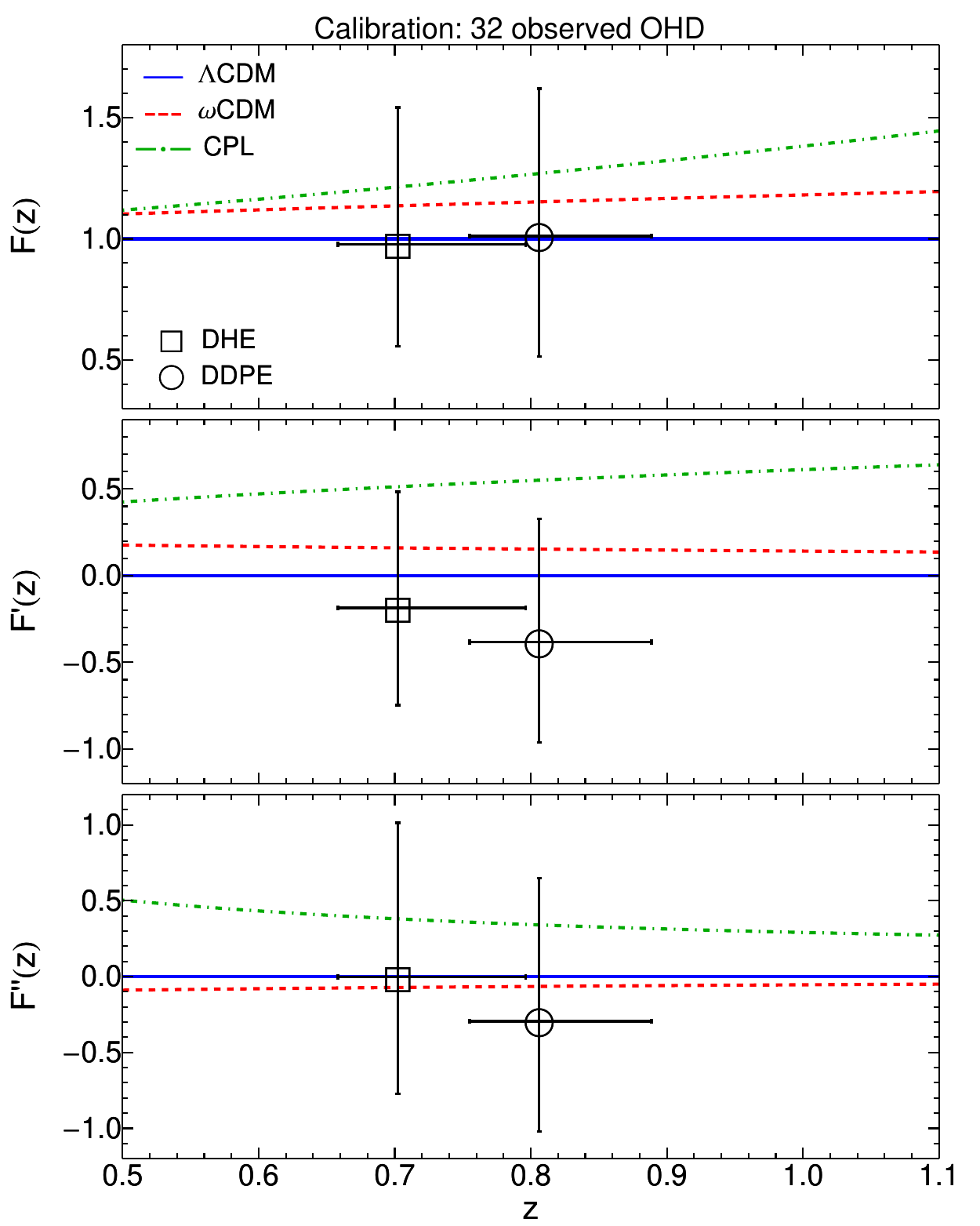}
\includegraphics[width=0.48\hsize,clip]{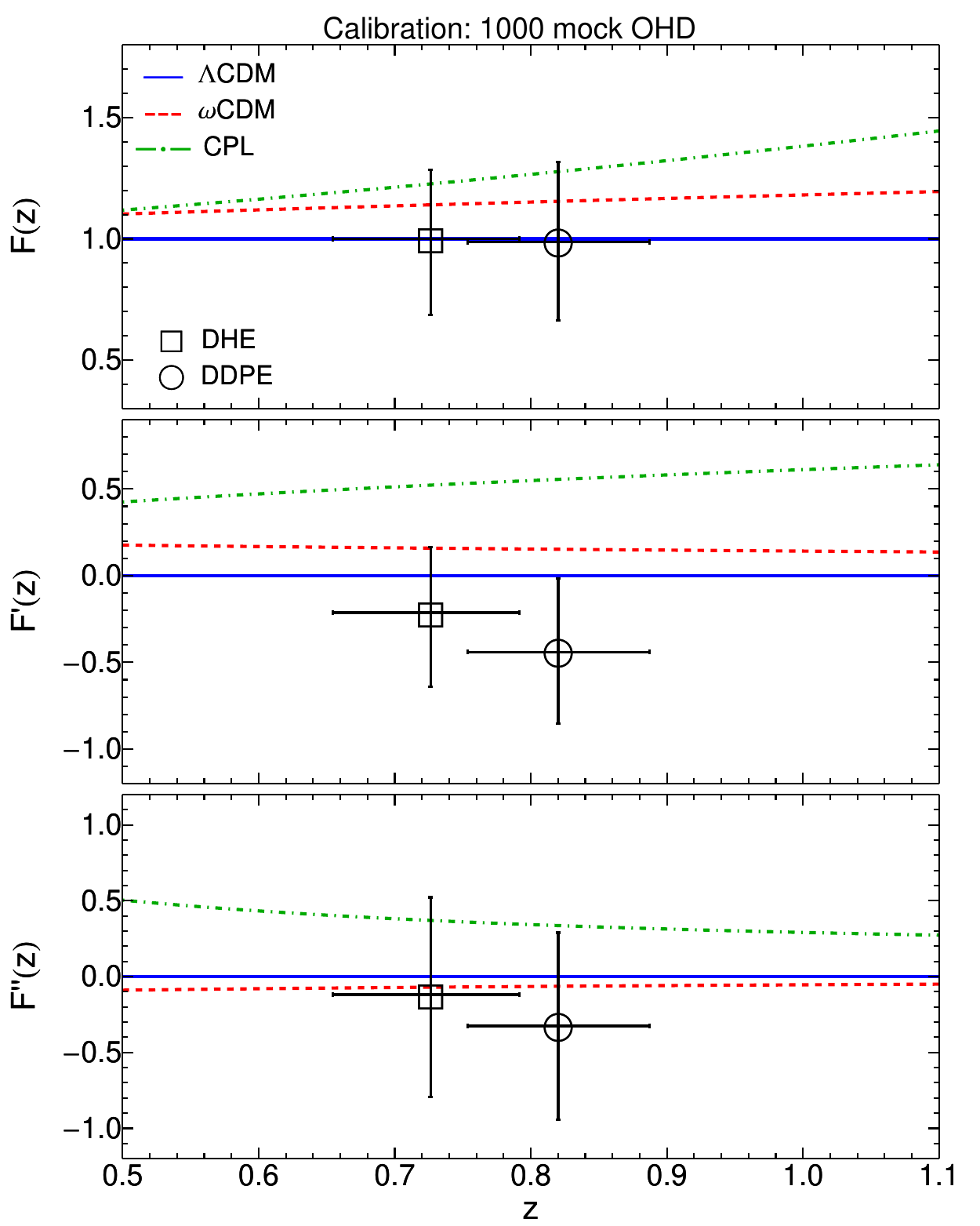}
\caption{Plots of $F(z)$, $F^\prime(z)$, and $F^{\prime\prime}(z)$ for $\Lambda$CDM, $\omega$CDM and CPL models, compared to the constraints on $F_{\rm tr}$, $F^\prime_{\rm tr}$, and $F^{\prime\prime}_{\rm tr}$ obtained from DHE and DDPE methods and calibrations based on the $32$ real and the $1000$ mock OHD measurements. Labels are shown in the legend of the upper panel.}
\label{fig:Fpl}
\end{figure*}

To better prompt this fact, we plot in Fig.~\ref{fig:Fpl} the behaviors of $F(z), F^\prime(z)$ and $F^{\prime\prime}(z)$ for the $\Lambda$CDM paradigm and the simplest extensions $\omega$CDM and CPL \citep{Chevallier2001,Linder2003}, with evolving dark energy, for which we have 
\begin{equation}
\left\{
    \begin{array}{lll}
    F(z)= 1 \,&,\quad & \hbox{$\Lambda$CDM}\\
    F(z)=(1+z)^{3(1+w_0)}\,&,\quad & \hbox{$\omega$CDM}\\
    F(z)=(1+z)^{3(1+w_0+w_1)}e^{-\frac{3w_1 z}{1+z}}\,&,\quad & \hbox{CPL}
        \end{array}
\right.\,.
\end{equation}
The parameter $w_0$ of the $\omega$CDM model and the parameters $w_0$ and $w_1$ of the CPL model are taken from \citet{2021PhRvD.104b3520B}.
Fig.~\ref{fig:Fpl} shows that at the level of $1$--$\sigma$ confidence level the DHE approach is more compatible than the standard cosmological model. Clearly the DDPE disfavors more the $\Lambda$CDM paradigm because it takes a direct expansion of $q(z)$ and then finds for the corresponding $H(z)$, leading to more complicated errors induced by the more complex structure of the method itself. 

Concluding, the two methods do not exclude dark energy to slightly evolve, albeit it is unlikely, especially from the DHE approach, that dark energy is not under the form of a  cosmological constant, implying that GRBs agree with the transition redshift predicted by the standard cosmological scenario.

\section{Conclusions and perspectives}\label{sezione5}

In this paper we derived model-independent constraints on the dark energy evolution through the evaluation of the transition redshift $z_{\rm tr}$.
The strategy we have pursued utilizes DHE and DDPE procedures introduced in \citet{2022MNRAS.509.5399C}, involving expansions of the Hubble rate $H(z)$ and the deceleration parameter $q(z)$, respectively. 
The constraints on $z_{\rm tr}$ have been obtained by using GRB \citep{2021JCAP...09..042K}, SNe Ia \citep{2018ApJ...853..126R} and uncorrelated BAO \citep{LM2020} catalogs.

The GRB data have been calibrated by resorting the well-known $E_{\rm p}$--$E_{\rm iso}$ correlation. This model-independent calibration involved the well-consolidate technique based on the B\'ezier polynomials interpolation of the OHD catalog \citep{2019MNRAS.486L..46A,LM2020}. 
In the current era of precision cosmology, we tested this calibration procedure by increasing the number of OHD points \citep[as outlined in Sec.~\ref{sezione2} and in][]{2022arXiv220513247K}. 
We found that the $1000$ mock OHD well reproduce the results obtained from the $32$ real measurements, of course with greater accuracy (see Sec.~\ref{sec:4}).
For comparison, we decided to calibrate GRBs using the real $32$ and the mock $1000$ OHD and to verify the stability of the correlation and cosmological parameters in different redshift bins containing roughly the same number of GRBs. 
The results from the joint calibrated GRBs~+~SNe~Ia~+~BAO MCMC fits, working out the Metropolis-Hastings algorithm and the DHE and the DDPE methods outlined in Sec.~\ref{sec:3}, are summarized in Tables ~\ref{tab:fits}--\ref{tab:zbins} and Figs.~\ref{fig:Bez_cont2}--\ref{fig:Bez_cont3}.
These results show an overall compatibility with respect to the previous literature and the $\Lambda$CDM paradigm.

The above MCMC fits show also that, although the $1000$ mock OHD data provide a more accurate reconstruction of $H(z)$, compared to that from the $32$ real measurements, but joint calibration and cosmological fits are insensitive to the OHD data sets employed to calibrate the $E_{\rm p}$--$E_{\rm iso}$ catalog.
This is reasonable because GRBs have the largest errors\footnote{GRB errors become even larger due to OHD uncertainties that propagate via $d_{\rm cal}(z)$.}, therefore, their weight in determining the best-fit results is certainly minor with respect to SNe Ia and BAO data.
However, the influence of OHD measurements (throught the calibration of GRB data) is evident when looking at the values of $h_0$ we have found from our MCMC fits (see Table~\ref{tab:fits}) which are consistent with those got from OHD (see $\alpha_0$ in Table~\ref{tab:Bez}) and in general from BAO measurements.

The advantage of using $1000$ mock OHD rather than the $32$ observed ones is more evident from the analysis performed in Sec.~\ref{sec:de_evol}.
Mainly by means of the more accurate reconstruction of $H(z)$ obtained by using the $1000$ mock OHD, inside the intervals of validity for $z_{tr}$, we notice that evolving dark energy is not fully excluded, albeit we underline that only small departures from the standard cosmological scenario are possible. In particular, we prompt the evolution of dark energy at the transition and we show the limits of evolving dark energy up to the transition time, adopting GRB data. 

Future developments will focus on constraining the transition epoch adopting more complicated combinations of data catalogs. We will also investigate more refined cosmographic expansions and techniques in order to feature the dark energy evolution at high redshifts.

\section*{Acknowledgements}
OL expresses his gratitude to the Instituto de Ciencias Nucleares of the UNAM University for  hospitality during the period in which this manuscript has been written. OL and MM acknowledge Alejandro Aviles, Alessandro Bravetti, Celia Escamilla-Rivera, Peter~K.S. Dunsby and Hernando Quevedo for fruitful discussions. The  work is financially supported by the Ministry of Education and Science of the Republic of Kazakhstan, Grant IRN AP08052311.  

\section*{Data Availability}
Data are available at the following references: SNe Ia catalog from \citet{2018ApJ...853..126R}, the updated $E_{\rm p}$--$E_{\rm iso}$ data set of $118$ long GRBs from \citet{2021JCAP...09..042K}, the uncorrelated BAO data from \citet{LM2020} and the updated $32$ OHD measurements from \citet{2022arXiv220513247K}.


\begin{thebibliography}{}
\makeatletter
\relax
\def\mn@urlcharsother{\let\do\@makeother \do\$\do\&\do\#\do\^\do\_\do\%\do\~}
\def\mn@doi{\begingroup\mn@urlcharsother \@ifnextchar [ {\mn@doi@}
  {\mn@doi@[]}}
\def\mn@doi@[#1]#2{\def\@tempa{#1}\ifx\@tempa\@empty \href
  {http://dx.doi.org/#2} {doi:#2}\else \href {http://dx.doi.org/#2} {#1}\fi
  \endgroup}
\def\mn@eprint#1#2{\mn@eprint@#1:#2::\@nil}
\def\mn@eprint@arXiv#1{\href {http://arxiv.org/abs/#1} {{\tt arXiv:#1}}}
\def\mn@eprint@dblp#1{\href {http://dblp.uni-trier.de/rec/bibtex/#1.xml}
  {dblp:#1}}
\def\mn@eprint@#1:#2:#3:#4\@nil{\def\@tempa {#1}\def\@tempb {#2}\def\@tempc
  {#3}\ifx \@tempc \@empty \let \@tempc \@tempb \let \@tempb \@tempa \fi \ifx
  \@tempb \@empty \def\@tempb {arXiv}\fi \@ifundefined
  {mn@eprint@\@tempb}{\@tempb:\@tempc}{\expandafter \expandafter \csname
  mn@eprint@\@tempb\endcsname \expandafter{\@tempc}}}

\bibitem[\protect\citeauthoryear{{Amati} \& {Della Valle}}{{Amati} \& {Della
  Valle}}{2013}]{AmatiDellaValle2013}
{Amati} L.,  {Della Valle} M.,  2013, \mn@doi [International Journal of Modern
  Physics D] {10.1142/S0218271813300280}, \href
  {http://adsabs.harvard.edu/abs/2013IJMPD..2230028A} {22, 1330028}

\bibitem[\protect\citeauthoryear{{Amati} et~al.,}{{Amati}
  et~al.}{2002}]{2002A&A...390...81A}
{Amati} L.,  et~al., 2002, \mn@doi [\aap] {10.1051/0004-6361:20020722}, \href
  {https://ui.adsabs.harvard.edu/abs/2002A&A...390...81A} {390, 81}

\bibitem[\protect\citeauthoryear{{Amati}, {D'Agostino}, {Luongo}, {Muccino}  \&
  {Tantalo}}{{Amati} et~al.}{2019}]{2019MNRAS.486L..46A}
{Amati} L.,  {D'Agostino} R.,  {Luongo} O.,  {Muccino} M.,   {Tantalo} M.,
  2019, \mn@doi [\mnras] {10.1093/mnrasl/slz056}, \href
  {https://ui.adsabs.harvard.edu/abs/2019MNRAS.486L..46A} {486, L46}

\bibitem[\protect\citeauthoryear{{Arjona}, {Cardona}  \& {Nesseris}}{{Arjona}
  et~al.}{2019}]{codice}
{Arjona} R.,  {Cardona} W.,   {Nesseris} S.,  2019, \mn@doi [\prd]
  {10.1103/PhysRevD.99.043516}, \href
  {https://ui.adsabs.harvard.edu/abs/2019PhRvD..99d3516A} {99, 043516}

\bibitem[\protect\citeauthoryear{Aviles, Gruber, Luongo  \& Quevedo}{Aviles
  et~al.}{2012a}]{Aviles:2012ay}
Aviles A.,  Gruber C.,  Luongo O.,   Quevedo H.,  2012a, \mn@doi [Phys. Rev. D]
  {10.1103/PhysRevD.86.123516}, 86, 123516

\bibitem[\protect\citeauthoryear{{Aviles}, {Gruber}, {Luongo}  \&
  {Quevedo}}{{Aviles} et~al.}{2012b}]{2012PhRvD..86l3516A}
{Aviles} A.,  {Gruber} C.,  {Luongo} O.,   {Quevedo} H.,  2012b, \mn@doi [\prd]
  {10.1103/PhysRevD.86.123516}, \href
  {https://ui.adsabs.harvard.edu/abs/2012PhRvD..86l3516A} {86, 123516}

\bibitem[\protect\citeauthoryear{Aviles, Bravetti, Capozziello  \&
  Luongo}{Aviles et~al.}{2014}]{Aviles:2014rma}
Aviles A.,  Bravetti A.,  Capozziello S.,   Luongo O.,  2014, \mn@doi [Phys.
  Rev. D] {10.1103/PhysRevD.90.043531}, 90, 043531

\bibitem[\protect\citeauthoryear{{Aviles}, {Klapp}  \& {Luongo}}{{Aviles}
  et~al.}{2017}]{2017PDU....17...25A}
{Aviles} A.,  {Klapp} J.,   {Luongo} O.,  2017, \mn@doi [Physics of the Dark
  Universe] {10.1016/j.dark.2017.07.002}, \href
  {https://ui.adsabs.harvard.edu/abs/2017PDU....17...25A} {17, 25}

\bibitem[\protect\citeauthoryear{{Banerjee}, {Cai}, {Heisenberg},
  {Colg{\'a}in}, {Sheikh-Jabbari}  \& {Yang}}{{Banerjee}
  et~al.}{2021}]{2021PhRvD.103h1305B}
{Banerjee} A.,  {Cai} H.,  {Heisenberg} L.,  {Colg{\'a}in} E.~{\'O}.,
  {Sheikh-Jabbari} M.~M.,   {Yang} T.,  2021, \mn@doi [\prd]
  {10.1103/PhysRevD.103.L081305}, \href
  {https://ui.adsabs.harvard.edu/abs/2021PhRvD.103h1305B} {103, L081305}

\bibitem[\protect\citeauthoryear{{Boshkayev}, {Konysbayev}, {Luongo}, {Muccino}
   \& {Pace}}{{Boshkayev} et~al.}{2021}]{2021PhRvD.104b3520B}
{Boshkayev} K.,  {Konysbayev} T.,  {Luongo} O.,  {Muccino} M.,   {Pace} F.,
  2021, \mn@doi [\prd] {10.1103/PhysRevD.104.023520}, \href
  {https://ui.adsabs.harvard.edu/abs/2021PhRvD.104b3520B} {104, 023520}

\bibitem[\protect\citeauthoryear{{Bull} et~al.,}{{Bull}
  et~al.}{2016}]{2016PDU....12...56B}
{Bull} P.,  et~al., 2016, \mn@doi [Physics of the Dark Universe]
  {10.1016/j.dark.2016.02.001}, \href
  {https://ui.adsabs.harvard.edu/abs/2016PDU....12...56B} {12, 56}

\bibitem[\protect\citeauthoryear{{Cao}, {Dainotti}  \& {Ratra}}{{Cao}
  et~al.}{2022}]{2022MNRAS.512..439C}
{Cao} S.,  {Dainotti} M.,   {Ratra} B.,  2022, \mn@doi [\mnras]
  {10.1093/mnras/stac517}, \href
  {https://ui.adsabs.harvard.edu/abs/2022MNRAS.512..439C} {512, 439}

\bibitem[\protect\citeauthoryear{Capozziello, Farooq, Luongo  \&
  Ratra}{Capozziello et~al.}{2014}]{Capozziello:2014zda}
Capozziello S.,  Farooq O.,  Luongo O.,   Ratra B.,  2014, \mn@doi [Phys. Rev.
  D] {10.1103/PhysRevD.90.044016}, 90, 044016

\bibitem[\protect\citeauthoryear{Capozziello, Luongo  \& Saridakis}{Capozziello
  et~al.}{2015}]{Capozziello:2015rda}
Capozziello S.,  Luongo O.,   Saridakis E.~N.,  2015, \mn@doi [Phys. Rev. D]
  {10.1103/PhysRevD.91.124037}, 91, 124037

\bibitem[\protect\citeauthoryear{Capozziello, D'Agostino  \&
  Luongo}{Capozziello et~al.}{2018a}]{Capozziello:2017nbu}
Capozziello S.,  D'Agostino R.,   Luongo O.,  2018a, \mn@doi [Mon. Not. Roy.
  Astron. Soc.] {10.1093/mnras/sty422}, 476, 3924

\bibitem[\protect\citeauthoryear{{Capozziello}, {D'Agostino}  \&
  {Luongo}}{{Capozziello} et~al.}{2018b}]{2018JCAP...05..008C}
{Capozziello} S.,  {D'Agostino} R.,   {Luongo} O.,  2018b, \mn@doi [\jcap]
  {10.1088/1475-7516/2018/05/008}, \href
  {https://ui.adsabs.harvard.edu/abs/2018JCAP...05..008C} {2018, 008}

\bibitem[\protect\citeauthoryear{Capozziello, D'Agostino  \&
  Luongo}{Capozziello et~al.}{2019}]{Capozziello:2019cav}
Capozziello S.,  D'Agostino R.,   Luongo O.,  2019, \mn@doi [Int. J. Mod. Phys.
  D] {10.1142/S0218271819300167}, 28, 1930016

\bibitem[\protect\citeauthoryear{Capozziello, D'Agostino  \&
  Luongo}{Capozziello et~al.}{2020}]{Capozziello:2020ctn}
Capozziello S.,  D'Agostino R.,   Luongo O.,  2020, \mn@doi [Mon. Not. Roy.
  Astron. Soc.] {10.1093/mnras/staa871}, 494, 2576

\bibitem[\protect\citeauthoryear{{Capozziello}, {Dunsby}  \&
  {Luongo}}{{Capozziello} et~al.}{2022}]{2022MNRAS.509.5399C}
{Capozziello} S.,  {Dunsby} P. K.~S.,   {Luongo} O.,  2022, \mn@doi [\mnras]
  {10.1093/mnras/stab3187}, \href
  {https://ui.adsabs.harvard.edu/abs/2022MNRAS.509.5399C} {509, 5399}

\bibitem[\protect\citeauthoryear{{Catto{\"e}n} \& {Visser}}{{Catto{\"e}n} \&
  {Visser}}{2007}]{2007CQGra..24.5985C}
{Catto{\"e}n} C.,  {Visser} M.,  2007, \mn@doi [Classical and Quantum Gravity]
  {10.1088/0264-9381/24/23/018}, \href
  {https://ui.adsabs.harvard.edu/abs/2007CQGra..24.5985C} {24, 5985}

\bibitem[\protect\citeauthoryear{{Catto{\"e}n} \& {Visser}}{{Catto{\"e}n} \&
  {Visser}}{2008a}]{2008CQGra..25p5013C}
{Catto{\"e}n} C.,  {Visser} M.,  2008a, \mn@doi [Classical and Quantum Gravity]
  {10.1088/0264-9381/25/16/165013}, \href
  {https://ui.adsabs.harvard.edu/abs/2008CQGra..25p5013C} {25, 165013}

\bibitem[\protect\citeauthoryear{{Catto{\"e}n} \& {Visser}}{{Catto{\"e}n} \&
  {Visser}}{2008b}]{2008PhRvD..78f3501C}
{Catto{\"e}n} C.,  {Visser} M.,  2008b, \mn@doi [\prd]
  {10.1103/PhysRevD.78.063501}, \href
  {https://ui.adsabs.harvard.edu/abs/2008PhRvD..78f3501C} {78, 063501}

\bibitem[\protect\citeauthoryear{{Chevallier} \& {Polarski}}{{Chevallier} \&
  {Polarski}}{2001}]{Chevallier2001}
{Chevallier} M.,  {Polarski} D.,  2001, \mn@doi [International Journal of
  Modern Physics D] {10.1142/S0218271801000822}, \href
  {http://adsabs.harvard.edu/abs/2001IJMPD..10..213C} {10, 213}

\bibitem[\protect\citeauthoryear{{Copeland}, {Sami}  \& {Tsujikawa}}{{Copeland}
  et~al.}{2006}]{reviewDE1}
{Copeland} E.~J.,  {Sami} M.,   {Tsujikawa} S.,  2006, \mn@doi [International
  Journal of Modern Physics D] {10.1142/S021827180600942X}, \href
  {https://ui.adsabs.harvard.edu/abs/2006IJMPD..15.1753C} {15, 1753}

\bibitem[\protect\citeauthoryear{{Cuceu}, {Farr}, {Lemos}  \&
  {Font-Ribera}}{{Cuceu} et~al.}{2019}]{2019JCAP...10..044C}
{Cuceu} A.,  {Farr} J.,  {Lemos} P.,   {Font-Ribera} A.,  2019, \mn@doi [\jcap]
  {10.1088/1475-7516/2019/10/044}, \href
  {https://ui.adsabs.harvard.edu/abs/2019JCAP...10..044C} {2019, 044}

\bibitem[\protect\citeauthoryear{{D'Agostini}}{{D'Agostini}}{2005}]{Dago2005}
{D'Agostini} G.,  2005, arXiv e-prints, \href
  {https://ui.adsabs.harvard.edu/abs/2005physics..11182D} {p. physics/0511182}

\bibitem[\protect\citeauthoryear{{Dainotti}, {Sarracino}  \&
  {Capozziello}}{{Dainotti} et~al.}{2022}]{2022PASJ..tmp...83D}
{Dainotti} M.~G.,  {Sarracino} G.,   {Capozziello} S.,  2022, \mn@doi [\pasj]
  {10.1093/pasj/psac057}, \href
  {https://ui.adsabs.harvard.edu/abs/2022PASJ..tmp...83D} {}

\bibitem[\protect\citeauthoryear{Demianski, Piedipalumbo, Sawant  \&
  Amati}{Demianski et~al.}{2021}]{Demianski2021}
Demianski M.,  Piedipalumbo E.,  Sawant D.,   Amati L.,  2021, \mn@doi [\mnras]
  {10.1093/mnras/stab1669}, 506, 903

\bibitem[\protect\citeauthoryear{{Di Valentino}, {Melchiorri}  \& {Silk}}{{Di
  Valentino} et~al.}{2020}]{2020NatAs...4..196D}
{Di Valentino} E.,  {Melchiorri} A.,   {Silk} J.,  2020, \mn@doi [Nature
  Astronomy] {10.1038/s41550-019-0906-9}, \href
  {https://ui.adsabs.harvard.edu/abs/2020NatAs...4..196D} {4, 196}

\bibitem[\protect\citeauthoryear{{Dunsby} \& {Luongo}}{{Dunsby} \&
  {Luongo}}{2016}]{2016IJGMM..1330002D}
{Dunsby} P. K.~S.,  {Luongo} O.,  2016, \mn@doi [International Journal of
  Geometric Methods in Modern Physics] {10.1142/S0219887816300026}, \href
  {https://ui.adsabs.harvard.edu/abs/2016IJGMM..1330002D} {13, 1630002}

\bibitem[\protect\citeauthoryear{Gruber \& Luongo}{Gruber \&
  Luongo}{2014}]{Gruber:2013wua}
Gruber C.,  Luongo O.,  2014, \mn@doi [Phys. Rev. D]
  {10.1103/PhysRevD.89.103506}, 89, 103506

\bibitem[\protect\citeauthoryear{{Handley}}{{Handley}}{2021}]{2021PhRvD.103d1301H}
{Handley} W.,  2021, \mn@doi ["Phys. Rev. D"] {10.1103/PhysRevD.103.L041301},
  \href {https://ui.adsabs.harvard.edu/abs/2021PhRvD.103d1301H} {103, L041301}

\bibitem[\protect\citeauthoryear{{Hawking} \& {Ellis}}{{Hawking} \&
  {Ellis}}{1973}]{1973lsss.book.....H}
{Hawking} S.~W.,  {Ellis} G.~F.~R.,  1973, {The large-scale structure of
  space-time.}

\bibitem[\protect\citeauthoryear{{Jia}, {Hu}, {Yang}, {Zhang}  \& {Wang}}{{Jia}
  et~al.}{2022}]{2022arXiv220809272J}
{Jia} X.~D.,  {Hu} J.~P.,  {Yang} J.,  {Zhang} B.~B.,   {Wang} F.~Y.,  2022,
  arXiv e-prints, \href {https://ui.adsabs.harvard.edu/abs/2022arXiv220809272J}
  {p. arXiv:2208.09272}

\bibitem[\protect\citeauthoryear{{Khadka}, {Luongo}, {Muccino}  \&
  {Ratra}}{{Khadka} et~al.}{2021}]{2021JCAP...09..042K}
{Khadka} N.,  {Luongo} O.,  {Muccino} M.,   {Ratra} B.,  2021, \mn@doi [\jcap]
  {10.1088/1475-7516/2021/09/042}, \href
  {https://ui.adsabs.harvard.edu/abs/2021JCAP...09..042K} {2021, 042}

\bibitem[\protect\citeauthoryear{{Khetan} et~al.,}{{Khetan}
  et~al.}{2021}]{2021A&A...647A..72K}
{Khetan} N.,  et~al., 2021, \mn@doi [\aap] {10.1051/0004-6361/202039196}, \href
  {https://ui.adsabs.harvard.edu/abs/2021A&A...647A..72K} {647, A72}

\bibitem[\protect\citeauthoryear{{King}, {Davis}, {Denney}, {Vestergaard}  \&
  {Watson}}{{King} et~al.}{2014}]{King2014}
{King} A.~L.,  {Davis} T.~M.,  {Denney} K.~D.,  {Vestergaard} M.,   {Watson}
  D.,  2014, \mn@doi [\mnras] {10.1093/mnras/stu793}, \href
  {http://adsabs.harvard.edu/abs/2014MNRAS.441.3454K} {441, 3454}

\bibitem[\protect\citeauthoryear{{Kumar}, {Jain}, {Mahajan}, {Mukherjee}  \&
  {Rana}}{{Kumar} et~al.}{2022}]{2022arXiv220513247K}
{Kumar} D.,  {Jain} D.,  {Mahajan} S.,  {Mukherjee} A.,   {Rana} A.,  2022,
  arXiv e-prints, \href {https://ui.adsabs.harvard.edu/abs/2022arXiv220513247K}
  {p. arXiv:2205.13247}

\bibitem[\protect\citeauthoryear{{Lee}, {Lee}, {{\'O} Colg{\'a}in},
  {Sheikh-Jabbari}  \& {Thakur}}{{Lee} et~al.}{2022}]{2022JCAP...04..004L}
{Lee} B.-H.,  {Lee} W.,  {{\'O} Colg{\'a}in} E.,  {Sheikh-Jabbari} M.~M.,
  {Thakur} S.,  2022, \mn@doi [\jcap] {10.1088/1475-7516/2022/04/004}, \href
  {https://ui.adsabs.harvard.edu/abs/2022JCAP...04..004L} {2022, 004}

\bibitem[\protect\citeauthoryear{{Liang}, {Xiao}, {Liu}  \& {Zhang}}{{Liang}
  et~al.}{2008}]{2008ApJ...685..354L}
{Liang} N.,  {Xiao} W.~K.,  {Liu} Y.,   {Zhang} S.~N.,  2008, \mn@doi [\apj]
  {10.1086/590903}, \href
  {https://ui.adsabs.harvard.edu/abs/2008ApJ...685..354L} {685, 354}

\bibitem[\protect\citeauthoryear{{Linder}}{{Linder}}{2003}]{Linder2003}
{Linder} E.~V.,  2003, \mn@doi [Physical Review Letters]
  {10.1103/PhysRevLett.90.091301}, \href
  {http://adsabs.harvard.edu/abs/2003PhRvL..90i1301L} {90, 091301}

\bibitem[\protect\citeauthoryear{Luongo}{Luongo}{2011}]{Luongo:2011zz}
Luongo O.,  2011, \mn@doi [Mod. Phys. Lett. A] {10.1142/S0217732311035894}, 26,
  1459

\bibitem[\protect\citeauthoryear{{Luongo} \& {Muccino}}{{Luongo} \&
  {Muccino}}{2018}]{luomuc}
{Luongo} O.,  {Muccino} M.,  2018, \mn@doi [\prd] {10.1103/PhysRevD.98.103520},
  \href {https://ui.adsabs.harvard.edu/abs/2018PhRvD..98j3520L} {98, 103520}

\bibitem[\protect\citeauthoryear{{Luongo} \& {Muccino}}{{Luongo} \&
  {Muccino}}{2021a}]{2021Galax...9...77L}
{Luongo} O.,  {Muccino} M.,  2021a, \mn@doi [Galaxies]
  {10.3390/galaxies9040077}, \href
  {https://ui.adsabs.harvard.edu/abs/2021Galax...9...77L} {9, 77}

\bibitem[\protect\citeauthoryear{{Luongo} \& {Muccino}}{{Luongo} \&
  {Muccino}}{2021b}]{LM2020}
{Luongo} O.,  {Muccino} M.,  2021b, \mn@doi [\mnras] {10.1093/mnras/stab795},
  \href {https://ui.adsabs.harvard.edu/abs/2021MNRAS.503.4581L} {503, 4581}

\bibitem[\protect\citeauthoryear{Luongo, Pisani  \& Troisi}{Luongo
  et~al.}{2016}]{Luongo:2015zgq}
Luongo O.,  Pisani G.~B.,   Troisi A.,  2016, \mn@doi [Int. J. Mod. Phys. D]
  {10.1142/S0218271817500158}, 26, 1750015

\bibitem[\protect\citeauthoryear{{Ma} \& {Zhang}}{{Ma} \&
  {Zhang}}{2011}]{2011ApJ...730...74M}
{Ma} C.,  {Zhang} T.-J.,  2011, \mn@doi [\apj] {10.1088/0004-637X/730/2/74},
  \href {https://ui.adsabs.harvard.edu/abs/2011ApJ...730...74M} {730, 74}

\bibitem[\protect\citeauthoryear{{Melchiorri}, {Pagano}  \&
  {Pandolfi}}{{Melchiorri} et~al.}{2007}]{2007PhRvD..76d1301M}
{Melchiorri} A.,  {Pagano} L.,   {Pandolfi} S.,  2007, \mn@doi [\prd]
  {10.1103/PhysRevD.76.041301}, \href
  {https://ui.adsabs.harvard.edu/abs/2007PhRvD..76d1301M} {76, 041301}

\bibitem[\protect\citeauthoryear{{Montiel}, {Cabrera}  \& {Hidalgo}}{{Montiel}
  et~al.}{2021}]{2021MNRAS.501.3515M}
{Montiel} A.,  {Cabrera} J.~I.,   {Hidalgo} J.~C.,  2021, \mn@doi [\mnras]
  {10.1093/mnras/staa3926}, \href
  {https://ui.adsabs.harvard.edu/abs/2021MNRAS.501.3515M} {501, 3515}

\bibitem[\protect\citeauthoryear{{Moresco}, {Jimenez}, {Verde}, {Cimatti}  \&
  {Pozzetti}}{{Moresco} et~al.}{2020}]{2020ApJ...898...82M}
{Moresco} M.,  {Jimenez} R.,  {Verde} L.,  {Cimatti} A.,   {Pozzetti} L.,
  2020, \mn@doi [\apj] {10.3847/1538-4357/ab9eb0}, \href
  {https://ui.adsabs.harvard.edu/abs/2020ApJ...898...82M} {898, 82}

\bibitem[\protect\citeauthoryear{{Moresco} et~al.,}{{Moresco}
  et~al.}{2022}]{2022LRR....25....6M}
{Moresco} M.,  et~al., 2022, \mn@doi [Living Reviews in Relativity]
  {10.1007/s41114-022-00040-z}, \href
  {https://ui.adsabs.harvard.edu/abs/2022LRR....25....6M} {25, 6}

\bibitem[\protect\citeauthoryear{{Muccino}, {Izzo}, {Luongo}, {Boshkayev},
  {Amati}, {Della Valle}, {Pisani}  \& {Zaninoni}}{{Muccino}
  et~al.}{2021}]{2021ApJ...908..181M}
{Muccino} M.,  {Izzo} L.,  {Luongo} O.,  {Boshkayev} K.,  {Amati} L.,  {Della
  Valle} M.,  {Pisani} G.~B.,   {Zaninoni} E.,  2021, \mn@doi [\apj]
  {10.3847/1538-4357/abd254}, \href
  {https://ui.adsabs.harvard.edu/abs/2021ApJ...908..181M} {908, 181}

\bibitem[\protect\citeauthoryear{{Ooba}, {Ratra}  \& {Sugiyama}}{{Ooba}
  et~al.}{2018}]{2018ApJ...864...80O}
{Ooba} J.,  {Ratra} B.,   {Sugiyama} N.,  2018, \mn@doi [\apj]
  {10.3847/1538-4357/aad633}, \href
  {https://ui.adsabs.harvard.edu/\#abs/2018ApJ...864...80O} {864, 80}

\bibitem[\protect\citeauthoryear{{Perivolaropoulos} \&
  {Skara}}{{Perivolaropoulos} \& {Skara}}{2022}]{2022NewAR..9501659P}
{Perivolaropoulos} L.,  {Skara} F.,  2022, \mn@doi [\nar]
  {10.1016/j.newar.2022.101659}, \href
  {https://ui.adsabs.harvard.edu/abs/2022NewAR..9501659P} {95, 101659}

\bibitem[\protect\citeauthoryear{{Planck Collaboration}}{{Planck
  Collaboration}}{2020}]{Planck2018}
{Planck Collaboration} 2020, \mn@doi [\aap] {10.1051/0004-6361/201833910},
  \href {https://ui.adsabs.harvard.edu/abs/2020A&A...641A...6P} {641, A6}

\bibitem[\protect\citeauthoryear{{Riess} et~al.,}{{Riess}
  et~al.}{2018}]{2018ApJ...853..126R}
{Riess} A.~G.,  et~al., 2018, \mn@doi [\apj] {10.3847/1538-4357/aaa5a9}, \href
  {https://ui.adsabs.harvard.edu/abs/2018ApJ...853..126R} {853, 126}

\bibitem[\protect\citeauthoryear{{Riess}, {Casertano}, {Yuan}, {Macri}  \&
  {Scolnic}}{{Riess} et~al.}{2019}]{2019ApJ...876...85R}
{Riess} A.~G.,  {Casertano} S.,  {Yuan} W.,  {Macri} L.~M.,   {Scolnic} D.,
  2019, \mn@doi [\apj] {10.3847/1538-4357/ab1422}, \href
  {https://ui.adsabs.harvard.edu/abs/2019ApJ...876...85R} {876, 85}

\bibitem[\protect\citeauthoryear{{Scolnic} et~al.,}{{Scolnic}
  et~al.}{2018}]{2018ApJ...859..101S}
{Scolnic} D.~M.,  et~al., 2018, \mn@doi [\apj] {10.3847/1538-4357/aab9bb},
  \href {https://ui.adsabs.harvard.edu/abs/2018ApJ...859..101S} {859, 101}

\bibitem[\protect\citeauthoryear{{Visser}}{{Visser}}{2005}]{2005GReGr..37.1541V}
{Visser} M.,  2005, \mn@doi [General Relativity and Gravitation]
  {10.1007/s10714-005-0134-8}, \href
  {https://ui.adsabs.harvard.edu/abs/2005GReGr..37.1541V} {37, 1541}

\bibitem[\protect\citeauthoryear{{Wang}, {Dai}  \& {Liang}}{{Wang}
  et~al.}{2015}]{2015NewAR..67....1W}
{Wang} F.~Y.,  {Dai} Z.~G.,   {Liang} E.~W.,  2015, \mn@doi [\nar]
  {10.1016/j.newar.2015.03.001}, \href
  {https://ui.adsabs.harvard.edu/abs/2015NewAR..67....1W} {67, 1}

\bibitem[\protect\citeauthoryear{{Weinberg}}{{Weinberg}}{1989}]{weinberg}
{Weinberg} S.,  1989, \mn@doi [Reviews of Modern Physics]
  {10.1103/RevModPhys.61.1}, \href
  {https://ui.adsabs.harvard.edu/abs/1989RvMP...61....1W} {61, 1}

\bibitem[\protect\citeauthoryear{{Yoo} \& {Watanabe}}{{Yoo} \&
  {Watanabe}}{2012}]{casc}
{Yoo} J.,  {Watanabe} Y.,  2012, \mn@doi [International Journal of Modern
  Physics D] {10.1142/S0218271812300029}, \href
  {https://ui.adsabs.harvard.edu/abs/2012IJMPD..2130002Y} {21, 1230002}

\makeatother
\end{thebibliography}
\end{document}